\documentclass[10pt]{article}

\usepackage{geometry}
\usepackage{hyperref}
\usepackage{multirow}
\usepackage{comment}
\usepackage{amsmath,amssymb}

\usepackage{algorithm}
\usepackage{graphicx}

\usepackage{theorem}
\newenvironment{definition}[1][Definition]{\begin{trivlist}
\item[\hskip \labelsep {\bfseries #1}]}{\end{trivlist}}

\newtheorem{theorem}{Theorem}[section]
\newtheorem{lemma}[theorem]{Lemma}
\newtheorem{assumption}{Assumption}[section]
\newtheorem{proposition}[theorem]{Proposition}

\newenvironment{example}[1][Example]{\begin{trivlist}
\item[\hskip \labelsep {\bfseries #1}]}{\end{trivlist}}

\newenvironment{proof}[1][Proof]{\begin{trivlist}
\item[\hskip \labelsep {\bfseries #1}]}{\end{trivlist}}

\usepackage{natbib}
 \bibpunct[, ]{(}{)}{,}{a}{}{,}%
\newcommand{\exclude}[1]{}

\def\Halmos{\mbox{\quad$\square$}}

\author{Jiajie Chen\\[4pt]{\small{\sc Department of Statistics, University of
    Wisconsin-Madison}}\\{\tt \small chen@stat.wisc.edu}\\[12pt]
Cong Han Lim\\[4pt]
{\sc \small Department of Computer Science, University of Wisconsin-Madison}\normalsize\\
{\tt \small conghan@cs.wisc.edu}\\[12pt]
Peter Z. G. Qian\\[4pt]{\small{\sc Department of Statistics, University of
    Wisconsin-Madison}}\\{\tt \small peterq@stat.wisc.edu}\\[12pt]
Jeff Linderoth\footnote{Work is partially supported through a contract from Argonne, a U.S.
Department of Energy Office of Science laboratory.  This work was supported by
the Applied Mathematics activity, Advance Scientific Computing Research program
within the DOE Office of Science}\\[4pt]
{\sc \small Department of Industrial and Systems Engineering, University of Wisconsin-Madison}\normalsize\\
{\tt \small linderoth@wisc.edu}\\[12pt]
Stephen J. Wright\footnotemark[1]\\[4pt]
{\sc \small Department of Computer Science, University of Wisconsin-Madison}\normalsize\\
{\tt \small swright@cs.wisc.edu}\\[12pt]
}

\title{Validating Sample Average Approximation Solutions with Negatively Dependent
Batches}

\exclude{

\documentclass[opre,nonblindrev]{informs3} 

\OneAndAHalfSpacedXI 

\usepackage{endnotes}
\let\footnote=\endnote

\TheoremsNumberedThrough     
\ECRepeatTheorems

\EquationsNumberedThrough    

}

\renewcommand{\vec}[1]{\mbox{\boldmath ${#1}$}}
\newcommand{\var}{{\rm var}}
\newcommand{\cov}{{\rm cov}}

\def\vD{D}

\def\0{\vec{0}}

\newcommand{\Rmnum}[1]{\expandafter\@slowromancap\romannumeral #1@}

\def\chlcomment#1{ }
\def\pqcomment#1{ }
\def\sjwcomment#1{ }
\def\jlcomment#1{ }
\def\jjcomment#1{ }

\begin{document}

\maketitle

\begin{abstract}
Sample-average approximations (SAA) are a practical means of
  finding approximate solutions of stochastic programming problems
  involving an extremely large (or infinite) number of scenarios. SAA
  can also be used to find estimates of a lower bound on the optimal
  objective value of the true problem which, when coupled with an
  upper bound, provides confidence intervals for the true optimal
  objective value and valuable information about the quality of the
  approximate solutions.  Specifically, the lower bound can be
  estimated by solving multiple SAA problems (each obtained using a
  particular sampling method) and averaging the obtained objective
  values. State-of-the-art methods for lower-bound estimation generate
  batches of scenarios for the SAA problems independently. In this
  paper, we describe sampling methods that produce negatively
  dependent batches, thus reducing the variance of the sample-averaged
  lower bound estimator and increasing its usefulness in defining a
  confidence interval for the optimal objective value. We provide
  conditions under which the new sampling methods can reduce the
  variance of the lower bound estimator, and present computational
  results to verify that our scheme can reduce the variance
  significantly, by comparison with the traditional Latin hypercube
  approach.
\end{abstract}

\exclude{

\RUNAUTHOR{Chen et al.} 

\RUNTITLE{Negatively Dependent Batches for SAA}

\TITLE{Validating Sample Average Approximation Solutions with Negatively Dependent
Batches}

\ARTICLEAUTHORS{%
\AUTHOR{Jiajie Chen}
\AFF{Department of Statistics, University of Wisconsin-Madison, \EMAIL{chen@stat.wisc.edu}} 
\AUTHOR{Cong Han Lim}
\AFF{Department of Computer Science, University of Wisconsin-Madison, \EMAIL{conghan@cs.wisc.edu}}
\AUTHOR{Peter Z. G. Qian}
\AFF{Department of Statistics, University of Wisconsin-Madison, \EMAIL{peterq@stat.wisc.edu}} 
\AUTHOR{Jeffrey T. Linderoth}
\AFF{Department of Industrial and Systems Engineering, University of Wisconsin-Madison, \EMAIL{linderoth@wisc.edu}}
\AUTHOR{Stephen J. Wright}
\AFF{Department of Computer Science, University of Wisconsin-Madison, \EMAIL{swright@cs.wisc.edu}} 
} 


\ABSTRACT{Sample-average approximations (SAA) are a practical means of
  finding approximate solutions of stochastic programming problems
  involving an extremely large (or infinite) number of scenarios. SAA
  can also be used to find estimates of a lower bound on the optimal
  objective value of the true problem which, when coupled with an
  upper bound, provides confidence intervals for the true optimal
  objective value and valuable information about the quality of the
  approximate solutions.  Specifically, the lower bound can be
  estimated by solving multiple SAA problems (each obtained using a
  particular sampling method) and averaging the obtained objective
  values. State-of-the-art methods for lower-bound estimation generate
  batches of scenarios for the SAA problems independently. In this
  paper, we describe sampling methods that produce negatively
  dependent batches, thus reducing the variance of the sample-averaged
  lower bound estimator and increasing its usefulness in defining a
  confidence interval for the optimal objective value. We provide
  conditions under which the new sampling methods can reduce the
  variance of the lower bound estimator, and present computational
  results to verify that our scheme can reduce the variance
  significantly, by comparison with the traditional Latin hypercube
  approach.

}%



\maketitle

%


}

\section{Introduction}\label{sec:intro}
Stochastic programming provides a means for formulating and solving
optimization problems that contain uncertainty in the model. When the
number of possible scenarios for the uncertainty is extremely large or
infinite, sample-average approximation (SAA) provides a means for
finding approximate solutions for a reasonable expenditure of
computational effort. In SAA, a finite number of scenarios is sampled
randomly from the full set of possible scenarios, and an approximation
to the full problem of reasonable size is formulated from the sampled
scenarios and solved using standard algorithms for deterministic
optimization (such as linear programming solvers).  Solutions obtained
from SAA procedures are typically suboptimal. Substantial research has
been done in assessing the quality of an obtained solution (or a set
of candidate solutions), and in understanding how different sampling
methods affect the quality of the approximate solution.

Important information about a stochastic optimization problem is
provided by the \emph{optimality gap} \citep{Mak1999,Bayraksan2006},
which provides a bound on the difference between the objective value
for the computed SAA solution and the true optimal objective value. An
estimate of the optimality gap can be computed using upper and lower
bounds on the true optimal objective value. \cite{Mak1999} proves
that the expected objective value of an SAA problem is a lower bound of
the objective of the true solution, and that this expected value
approaches the true optimal objective value as the number of scenarios
increases. We can estimate this lower bound (together with confidence
intervals) by solving multiple SAA problems, a task that can be
implemented in parallel in an obvious way.  An upper bound can be
computed by taking a candidate solution $x$ and evaluating the
objective by sampling from the scenario set, typically taking a larger
number of samples than were used to set up the SAA optimization
problem for computing $x$.

Much work has been done to understand the quality of SAA solutions
obtained from Monte Carlo (MC) sampling, Latin hypercube (LH) sampling
\citep{Mckay1979}, and other methods. MC generates independent
identically distributed scenarios where the value of each variable is
picked independently from its corresponding distribution. The
simplicity of this method has made it an important practical tool; it
has been the subject of much theoretical and empirical research. Many
results about the asymptotic behavior of optimal solutions and values
of MC have been obtained; see \cite[Chapter~5]{Shapiro2009} for a
review. By contrast with MC, LH stratifies each dimension of the
sample space in such a way that each strata has the same probability, then samples the scenarios so that each strata is represented in the
scenario sample set. This procedure introduces a dependence between
the different scenarios of an individual SAA problem. The sample space
is in some sense ``uniformly'' covered on a per-variable basis, thus
ensuring that there are no large unsampled regions and leading to
improved performance. \cite{Linderoth2006} provides empirical results
showing that the bias and variance of a lower bound obtained by
solving multiple SAA problems constructed with LH sampling is
considerably smaller than the statistics obtained from an MC-based
procedure.  Theoretical results about the asymptotic behavior of these
estimates were provided later by \cite{Mello2008}.  Other results on
the performance of LH have been obtained, including results on large
deviations \citep{Drew2005}, rate of convergence to optimal values
\citep{Mello2008}, and bias reduction of approximate optimal values
\citep{Freimer2012}, all of which demonstrate the
superiority of LH over MC. This favorable empirical and theoretical
evidence makes LH the current state-of-the-art method for obtaining
high-quality lower bounds and optimality gaps via SAA. In this paper,
we build on the ideas behind the LH method to obtain LH variants with
even better variance properties.

In the past, when solving a set of SAA problems to obtain a
lower-bound estimate of the true optimal objective value, each batch
of scenarios determining each SAA was chosen independently of the
other batches.
In this paper, we introduce two approaches to sampling --- {\em sliced
  Latin hypercube (SLH)} sampling and {\em sliced orthogonal-array
  Latin hypercube (SOLH)} sampling --- that yield better estimates of
the lower bound by imposing negative dependencies between the batches
in the different SAA approximations. 
  These approaches not only stratify {\em within} each batch
  (as in LH) but also {\em between
  all batches}. 
The SLH approach is easy to implement, while the SOLH approach 
provides better variance reduction. With these methods, we can 
significantly reduce the variance of the lower bound estimator even 
if the size of each SAA problem or the number of SAA problems were 
kept the same, which can be especially useful when solving each SAA 
problem is time consuming or when computing resources are limited. 
We will provide theoretical results analyzing the variance reduction 
properties of both approaches, and present empirical results 
demonstrating their efficacy across a variety of stochastic programs 
studied in the literature. Sliced Latin hypercube sampling was 
introduced first in \cite{Qian2012} and has proven to be useful for 
the collective evaluation of computer models and ensembles of computer 
models.

Here we briefly outline the rest of our paper.  The next section
begins with a brief description of how the optimality gap can be
estimated (\S~\ref{subsec:lbe}), a review of Latin hypercube sampling
(\S~\ref{subsec:lhd}) and an introduction of functional analysis of
variance (\S~\ref{subsec:anova}). Section~\ref{sec:slhd} focuses on
sliced Latin hypercube sampling.  It outlines the construction of
dependent batches (\S~\ref{subsec:slhd_con}), describes theoretical
results of variance reduction based on a certain monotonicity
condition (\S~\ref{subsec:samp_slhd}), applies the results to
two-stage stochastic linear program (\S~\ref{subsec:2stage}), and
finally studies the relation between the lower bound estimation and
numerical integration (\S~\ref{subsec:link}). Section~\ref{sec:oslhd}
reviews orthogonal arrays and introduces a method to incorporate these
arrays into the sliced Latin hypercube sampling, which leads to
stronger between-batch negative dependence. The next two sections deal
with our computational experiments. Section~\ref{sec:expsetup}
describes the setup and some of the implementation details, while
Section~\ref{sec:compute} describes and analyzes the performance of
the new sampling methods in the lower bound estimation problem. We end
the paper in Section~\ref{sec:dis} with a summary of our results and a discussion of possible
future research.

\section{Preliminaries}\label{sec:prelim}

\subsection{Stochastic Programs and Lower Bound Estimators}\label{subsec:lbe}
We consider a stochastic program of the form
\begin{equation}\label{eq:sp}
    \min_{\substack{x \in X}} \: g(x):=\mathbb{E}[G(x,\xi)],
\end{equation}
where $X \subset \mathbb{R}^p$ is a compact feasible set, $x \in X$ is
a vector of decision variables, $\xi=(\xi^1,\xi^2,\dotsc,\xi^m)$ is a
vector of random variables, and $G: \mathbb{R}^p \times \mathbb{R}^m
\rightarrow \mathbb{R}$ is a real-valued measurable function. Unless
stated otherwise, we assume that $\xi$ is a random vector with uniform
distribution on $(0,1]^m$ and that $\mathbb{E}$ is the expectation
  with respect to the distribution of $\xi$.  If $\xi$ has a different
  distribution on $\mathbb{R}^m$, we can transform it into a uniform
  random vector on $(0,1]^m$ as long as $\xi^1,\xi^2,\dotsc,\xi^m$ are
    either (i) independent discrete or continuous random variables or
    (ii) dependent random variables which are absolutely continuous
    \citep{Rosenblatt1953}.

Problem \eqref{eq:sp} may be extremely challenging to solve directly,
since the evaluation of $g(x)$ involves a high-dimensional
integration.  We can replace \eqref{eq:sp} with the
following approximation: 
\begin{equation}\label{eq:spn}
    \min_{\substack{x \in X}} \: \hat{g}_n(x):=\frac{1}{n}\sum_{i=1}^n
    G(x,\xi_i), 
\end{equation} 
where $\xi_1, \xi_2, \dotsc, \xi_n$ are scenarios sampled from the uniform
distribution on $(0,1]^m$. The function $\hat{g}_n$ is called a {\em
sample-average approximation (SAA)} to the objective $g$ in
\eqref{eq:sp}. In this paper we will frequently use the term \emph{SAA
problem} to refer to equation \eqref{eq:spn}. We use $x^*$ and $v^*$ to
denote a solution and the optimal value of the true problem
(\ref{eq:sp}), respectively, while $x_n^*$ and $v_n^*$ denote the
solution and optimal value of the SAA problem \eqref{eq:spn},
respectively.

We introduce here some items of terminology that are used throughout
the remainder of the paper. Let $D$ denote an $n \times m$ matrix
with $\xi_i^T$ in \eqref{eq:spn} as its $i$th row. Hence, $D$
represents a batch of scenarios that define an SAA problem. We will
refer $\xi_i$ to as the $i$th scenario in $D$. We use $D(:,k)$ to denote
the $k$th column of $\vD$, and $\xi_{ik}$ to denote the $(i,k)$ entry
of $D$, that is, the $k$th entry in the $i$th scenario in $D$.

We can express the dependence of $v_n^*$ in \eqref{eq:spn} on $D$
explicitly by writing this quantity as $v_n(D)$, where $v_n: (0,1]^{n
\times m} \rightarrow \mathbb{R}$.  Given a distribution over the $D$
matrices where $\xi_1,\xi_2,\dotsc,\xi_n$ are each drawn from the 
uniform $(0,1]^m$ distribution but not necessarily independently, it is well
known and easy to show that the expectation with respect to the $D$
matrices $\mathbb{E}[v_n(D)] \leq v^*$ giving us a lower
bound of the true optimal value
\citep{norkin.pflug.ruszczynski:98,Mak1999}. $\mathbb{E}[v_n(D)]$ can be
estimated as follows. Generate $t$ independent batches
$D_1,D_2,\dotsc,D_t$ and compute the optimal value $\min \;v_n(D_r)$ by
solving \emph{subproblem} \eqref{eq:spn} for each $\vD_r$,
$r=1,2,\dotsc,t$. From \cite{Mak1999}, a lower bound estimate of $v^*$
is
\begin{equation} \label{eq:lbe}
    L_{n,t}:=\frac{v_n(D_1)+v_2(D_2)+\cdots+v_n(D_t)}{t}.
\end{equation}

\subsection{Latin Hypercube Sampling}\label{subsec:lhd}

Latin hypercube sampling, which stratifies sample points along each
dimension \citep{Mckay1979}, has been used in numerical integration
for many years. It has been shown that the variance of the mean output
of a computer experiment under Latin hypercube sampling can be much
lower than for experiments based on Monte Carlo methods
\citep{Mckay1979,Stein1987,Loh1996}.  Let $v_n^{MC}(D)$ and
$v_n^{LH}(D)$ denote the approximate optimal value when the $\xi_i$ in
$D$ are generated using Monte Carlo and Latin hypercube sampling,
respectively. \cite{Mello2008} showed that the asymptotic variance of
$v_n^{LH}(D)$ is smaller than the variance of $v_n^{MC}(D)$ under some
fairly general conditions. Extensive numerical simulations have
provided empirical demonstrations of the superiority of Latin
hypercube sampling \citep{Mello2008,Linderoth2006}.

To define Latin hypercube sampling, we start with some useful
notation. Given an integer $p \geq 1$, we define $Z_p :=
\{1,2,\dotsc,p\}$. Given an integer $a$, the notation $Z_p \oplus a$
denotes the set $\{a + 1, a + 2, \dotsc , a + p\}$. For a real number
$y$, $\left\lceil y \right\rceil$ denotes the smallest integer no less
than $y$. A ``uniform permutation on a set of $p$ integers'' is
obtained by randomly taking a permutation on the set, with all $p!$
permutations being equally probable. 

We have the following definition.
\begin{definition}\label{def:olh}
An $n \times m$ array $A$ is a {\em Latin hypercube} if each column of
$A$ is a uniform permutation on $Z_n$. Moreover, $A$ is an {\em
  ordinary Latin hypercube} if all its columns are generated
independently.
\end{definition}
Using an ordinary Latin hypercube $A$, an $n \times m$ matrix $\vD$
with scenarios $\xi_1,\xi_2,\dotsc,\xi_n$ that defines an SAA problem
is obtained as follows \citep{Mckay1979}:
\begin{equation}\label{eq:lhsc}
    \xi_{ik}=\frac{a_{ik}-\gamma_{ik}}{n}, \quad i=1,2,\dotsc,n, \quad
    k=1,2,\dotsc,m,
\end{equation}
where all $\gamma_{ik}$ are $U[0,1)$ random variables, and the
  quantities $a_{ik}$ and the $\gamma_{ik}$ are mutually
  independent. We refer the matrix $D$ thus constructed as an
  \emph{ordinary Latin hypercube design}. 

We now introduce a different way of looking at design matrices $D\in
(0,1]^{n \times m}$ that will be useful when we discuss extensions to
sliced Latin hypercube designs in later sections. We can write a design
matrix $D$ as
\begin{equation}\label{eq:A-D}
    D=(B-\Theta)/n,
\end{equation}
where 
\begin{align*}
B &=(b_{ik})_{n\times m}, \quad \mbox{with $b_{ik}=\left\lceil n \xi_{ik}
\right\rceil$}, \\
\Theta &=(\theta_{ik})_{n \times m}, \quad \mbox{with
$\theta_{ik}=b_{ik}-n\xi_{ik}$.}
\end{align*}
When $D$ is an ordinary Latin hypercube design, $B$ is an ordinary Latin
hypercube and $\theta_{ik}$ corresponds to $\gamma_{ik}$ in
\eqref{eq:lhsc} for all $i=1,2,\dotsc,n$ and $k=1,2,\dotsc,m$. By the
properties of an ordinary Latin hypercube design, the entries in
$\Theta$ are mutually independent, and $\Theta$ and $B$ are independent.
We refer $B$ to as the \emph{underlying array} of $D$. 

The lower bound on $v^*$ can be estimated from \eqref{eq:lbe} by
taking $t$ {\em independently generated} ordinary Latin hypercube
designs $D_1,D_2,\dotsc,D_t$ \citep{Linderoth2006}. We denote this
sampling scheme by ILH and denote the estimator obtained from
\eqref{eq:lbe} by $L_{n,t}^{ILH}$.

\begin{figure}[htbp]
\begin{center}
\includegraphics[scale=1]{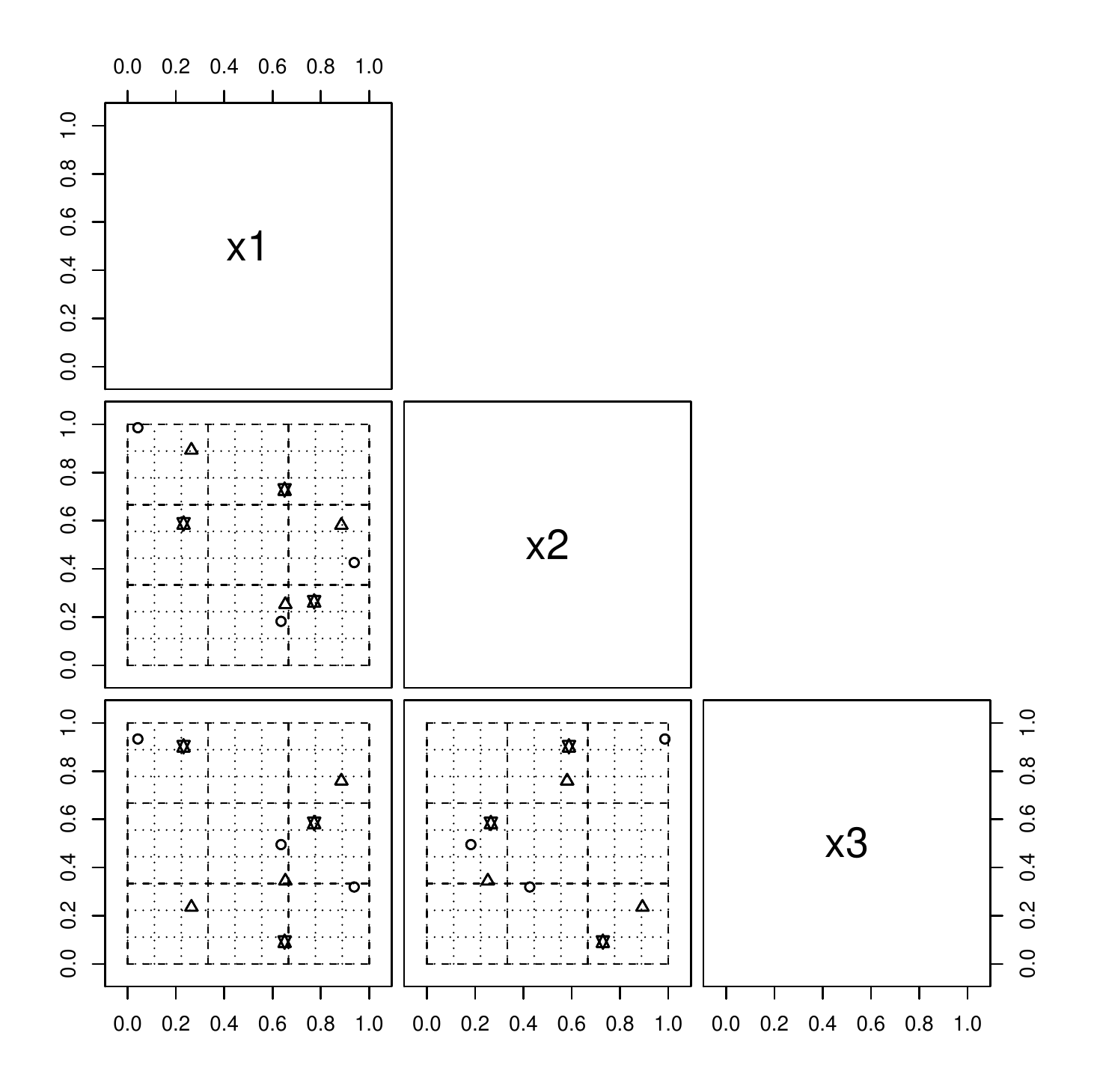}
\caption{Bivariate projections of three independent $3 \times 3$ ordinary Latin hypercube designs.}
\label{fig:ilhd}
\end{center}
\end{figure}

 To illustrate the limitations of the ILH scheme,
 Figure~\ref{fig:ilhd} displays three independent $3 \times 3$
 ordinary Latin hypercube designs generated under ILH with $n=t=m=3$.
 Scenarios from each three-dimensional design are denoted by the same
 symbol, and are projected onto each of the three bivariate
 planes. The dashed lines stratify each dimension into three
 partitions. For any design, each of the these three intervals will
 contain exactly one scenario in each dimension. This scheme covers the
 space more ``uniformly'' than three scenarios that are picked identically
 and independently from the uniform distribution, as happens in Monte
 Carlo schemes.  However, the combination of points from all three
 designs does not cover the space particularly well, which gives some
 cause for concern, since all designs are being used in the lower
 bound estimation. Specifically, when we combine the three designs
 together (to give nine scenarios in total), it is usually {\em not} the
 case that each of the nine equally spaced intervals of (0,1] contains
exactly one scenario in any dimension.  This shortcoming provides the
intuition behind the sliced Latin hypercube (SLH) design, which we
will describe in the subsequent sections.

\subsection{Functional ANOVA Decomposition}\label{subsec:anova}
In order to understand the asympototic properties of estimators
arising from Latin-Hypercube based samples, it is necessary to 
review the functional analysis of variance decomposition
\citep{Owen1994}, also known as {\em functional ANOVA}.  Let
$\mathcal{D}:=\{1,2,\dotsc,m\}$ represent the axes of $(0,1]^m$
  associated with an input vector $\xi=(\xi^1,\xi^2,\dotsc,\xi^m)$
  defined in Section~\ref{subsec:lbe}.  Let $F_k$ denote the uniform
  distribution for $\xi^k$, with $F:=\prod_{k=1}^m F_k$. For $u
  \subseteq \mathcal{D}$, let $\xi^u$ denote a vector consisting of
  $\xi^{k}$ for $k \in u$.  Define
\begin{equation}\label{eq:anova1}
f_u(\xi^u):=\int \{f(\xi)-\sum_{v \subset u} f_v(\xi^v)\}dF_{\mathcal{D}-u},
\end{equation}
where $dF_{\mathcal{D}-u}=\prod_{k \notin u} dF_k$ integrates out all
components except for those included in $u$, and $v \subset u$ is a
proper subset of $u$. Hence, we have that 
\begin{itemize}
\item $f_{\emptyset}(\xi^\emptyset)=\int f(\xi)dF$ is the mean of
  $f(\xi)$;
\item
  $f_{\{k\}}(\xi^k)=\int\{f(\xi)-f_{\emptyset}(\xi^\emptyset)\}dF_{\mathcal{D}-\{k\}}$
  is the main effect function for factor $k$, and
\item
  $f_{\{k,l\}}(\xi^k,\xi^l)=\int\{f(\xi)-f_{\{k\}}(\xi^k)-f_{\{l\}}(\xi^l)-f_{\emptyset}(\xi^\emptyset)\}dF_{\mathcal{D}-\{k,l\}}$
  is the bivariate interaction function between factors $k$ and $l$.
\end{itemize}

When the stochastic program \eqref{eq:sp} has a unique solution $x^*$
and some mild conditions are satisfied, one has
\begin{equation*}
\frac{v_n^{MC}(D)-v^*}{\sigma_{n,MC}(x^*)} \xrightarrow{d}
\text{Normal}(0,1),
\end{equation*}
where $\sigma_{n,MC}^2 :=n^{-1}\var[G(x^*,\xi)]$ \citep{Shapiro1991}.
With additional assumptions, \cite{Mello2008} shows that 
\begin{equation}
\frac{v_n^{LH}(D)-v^*}{\sigma_{n,LH}(x^*)}
\xrightarrow{d} \text{Normal}(0,1),
\end{equation}
where $\sigma_{n,LH}^2 :=n^{-1}\var[G(x^*,\xi)]-n^{-1} \sum_{k=1}^m
\var[G_{\{k\}}(x^*,\xi^k)]+o(n^{-1})$ and $G_{\{k\}}(x^*,\xi^k)$ is the main effect function of $G(x^*,\xi)$
as defined in \eqref{eq:anova1}.

\section{Sliced Latin Hypercube Sampling}\label{sec:slhd}

Instead of generating $D_1,D_2,\dotsc,D_t$ independently for each SAA
subproblem, we propose a new scheme called {\em sliced Latin hypercube
  (SLH) sampling} that generates a family of correlated designs. An
SLH design \citep{Qian2012} is a $nt \times m$ matrix that can be
partitioned into $t$ separate LH designs, represented by the matrices
$D_r$, $r=1,2,\dotsc,t$, each having the same properties as ILH, with
respect to SAA.  SLH was originally introduced to aid in the 
collective evaluation of computer models, but here we demonstrate its
utility in creating multiple SAA problems to solve.

Let $L_{n,t}^{SLH}$ denote the lower bound estimator of $v^*$ under
SLH.
Because the individual designs $D_r$, $r=1,2,\dotsc,t$ are LH designs,
we have that
$\mathbb{E}(L_{n,t}^{SLH})=\mathbb{E}(L_{n,t}^{ILH})$. Consider two
distinct batches of scenarios $D_r$ and $D_s$ for any
$r,s=1,2,\dotsc,t$ and $r \neq s$. We will show that when $v_n(D)$
fulfills a certain monotonicity condition, $v_n(D_r)$ and $v_n(D_s)$
are negatively dependent under SLH. Compared with ILH, SLH
introduces \emph{between-batch} negative dependence while keeping the
\emph{within-batch} structure intact. As a result, we expect a
lower-variance estimator: $\var(L_{n,t}^{SLH}) \leq
\var(L_{n,t}^{ILH})$.

\subsection{Construction}\label{subsec:slhd_con}

Algorithm~\ref{alg:slhd} describes the construction of the matrices
$D_r$, $r=1,2,\dotsc,t$ for the SLH design. We use notation
$D_{r}(:,k)$ for the $k$th column of $D_r$, $\xi_{r,i}$ for the $i$th
scenario of $D_r$, and $\xi_{r,ik}$ for the $k$th entry in $\xi_{r,i}$, for $i=1,2,\dotsc,n$, $k=1,2,\dotsc,m$, and $r=1,2,\dotsc,t$.

\begin{algorithm}
\caption{Generating a Sliced Latin Hypercube Design}
\label{alg:slhd}
\begin{itemize}
\item[] {\bf Step 1.}  Randomly generate $t$ independent ordinary Latin
  hypercubes $A_r = (a_{r,ik})_{n\times m}$, $r=1,2,\dotsc,t$.
Denote the $k$th column of $A_r$ by $A_{r}(:,k)$, for $k=1,\ldots,m$.
\item[] {\bf Step 2.}  For $k=1,2,\dotsc,m$, do the following
  independently: Replace all the $\ell$s in $A_1(:,k),A_2(:,k),
  \dotsc,A_t(:,k)$ by a random permutation on $Z_t \oplus t(\ell-1)$,
  for $\ell=1,2,\dotsc,n$.
\item[] {\bf Step 3.}  For $r=1,2,\dotsc,t$, obtain the $(i,k)$th entry
  of $D_r$ as follows:
\begin{equation}\label{eq:slhd3}
\xi_{r,ik}=\frac{a_{r,ik} - \gamma_{r,ik}}{nt},
\end{equation}
where the $\gamma_{r,ik}$ are $U[0,1)$ random variables that are
  mutually independent of the $a_{r,ik}$.
\end{itemize}
\end{algorithm}
By vertically stacking the matrices $D_1,D_2,\dotsc,D_t$, we obtain
the $nt \times m$ matrix representing the SLH design, as defined in
\cite{Qian2012}.

As in \eqref{eq:A-D}, we can express each $D_r$ as
\begin{equation}\label{eq:B-D}
D_r=(B_r-\Theta_r)/n,
\end{equation}
where $B_r=(b_{r,ik})_{n\times m}$ with $b_{r,ik}=\left\lceil n
\xi_{r,ik} \right\rceil$ and $\Theta_r=(\theta_{r,ik})_{n \times m}$.
We have the following proposition regarding properties of the SLH
design, including dependence of the batches.  (This result is closely
related to \cite[Lemma~2]{Qian2012}.)
\begin{proposition}\label{prop:sigma}
Consider the SLH design with $D_r$, $r=1,2,\dotsc,t$ constructed according
to Algorithm~\ref{alg:slhd}, with $B_r$ and $\Theta_r$, $r=1,2,\dotsc,t$
defined as in \eqref{eq:B-D}. 
The following are true.
\begin{enumerate}
\item[(i)] $B_r$, $r=1,2,\dotsc,t$ are independent ordinary Latin
  hypercubes.
\item[(ii)] $B_r$ and $\Theta_r$ are independent, for each
  $r=1,2,\dotsc,t$.
\item[(iii)] Within each $\Theta_r$, $r=1,2,\dotsc,t$, the
  $\theta_{r,ik}$ are mutually independent $U[0,1)$ random variables.
\item[(iv)] For $r, s = 1,2,\dotsc,t$ with $r \neq s$, $\theta_{r,ik}$
  is dependent on $\theta_{s,ik}$ if and only if $B_{r,ik} =
  B_{s,ik}$;
\item[(v)] The $D_r$, $r=1,2,\dotsc,t$ are ordinary Latin hypercube
  designs, but they are not independent.
\end{enumerate}
\end{proposition}
\proof
\begin{enumerate}
\item [(i)] According to \eqref{eq:B-D} and Step 2 in the above
  construction, $B_r$ is the same as $A_r$ in Step 1 {\em prior to}
  the replacement step, and the result follows.
\item [(ii)] Note that $b_{r,ik}=\left\lceil
  \frac{a_{r,ik}-\gamma_{r,ik}}{t}\right\rceil=\left\lceil
  \frac{a_{r,ik}}{t}\right\rceil$ where the $a_{r,ik}$ are values in
  $A_r$ after the replacement in Step 2 of the construction above. By
  \eqref{eq:slhd3} and \eqref{eq:B-D}, we have
\begin{equation}\label{eq:sigma}
\theta_{r,ik}=b_{r,ik}-\frac{a_{r,ik}}{t}+\frac{\gamma_{r,ik}}{t}=\left(t\left\lceil
\frac{a_{r,ik}}{t}\right\rceil-a_{r,ik} \right)/t+\gamma_{r,ik}/t.
\end{equation}
According to Step 2 of the construction, for each $r=1,2,\dotsc,t$,
$i=1,2,\dotsc,n$ and $k=1,2,\dotsc,m$, the quantity $t\lceil
\frac{a_{r,ik}}{t}\rceil-a_{r,ik}$ is randomly selected among $Z_t$
and is independent of $B_r$. Since the $\gamma_{r,ik}$ are independent
of the $a_{r,ik}$, the claim is proved.
\item [(iii)] For each $r=1,\ldots,t$ and $k=1,\ldots,m$, the
  quantities $\{t\left\lceil \frac{a_{r,1k}}{t}\right\rceil-a_{r,1k},
  t\left\lceil \frac{a_{r,2k}}{t}\right\rceil-a_{r,2k},
  \dotsc,t\left\lceil \frac{a_{r,nk}}{t}\right\rceil-a_{r,nk}\}$ are
  independently and randomly selected among $n$ different $Z_t$'s,
  respectively. Thus, the $\theta_{r,ik}$ are mutually independent
  within $\Theta_r$. In other words, the $t\lceil
  \frac{a_{r,ik}}{t}\rceil-a_{r,ik}$, $i=1,2,\dotsc,n$ are mutually
  independent discrete uniform random variables on $Z_t$, such that
  the $\theta_{r,ik}$ are $U[0,1)$ random variables, by
    \eqref{eq:sigma}.
\item [(iv)] It suffices to show that $t\left\lceil
  \frac{a_{r,1k}}{t}\right\rceil-a_{r,1k}$ and $t\left\lceil
  \frac{a_{r,nk}}{t}\right\rceil-a_{s,nk}$ are dependent if and only
  if $B_{r,ik} = B_{s,ik}$. That is true because $t\left\lceil
  \frac{a_{r,1k}}{t}\right\rceil-a_{r,1k}$ and $t\left\lceil
  \frac{a_{r,nk}}{t}\right\rceil-a_{s,nk}$ are selected from the same
  $Z_t$ when $B_{r,ik} = B_{s,ik}$.
\item [(v)] The result follows directly from (i), (ii), (iv), and the
  definition of the ordinary Latin hypercube design. \Halmos
\end{enumerate}
\endproof

\begin{figure}[htbp]
\centering
\includegraphics[scale=1]{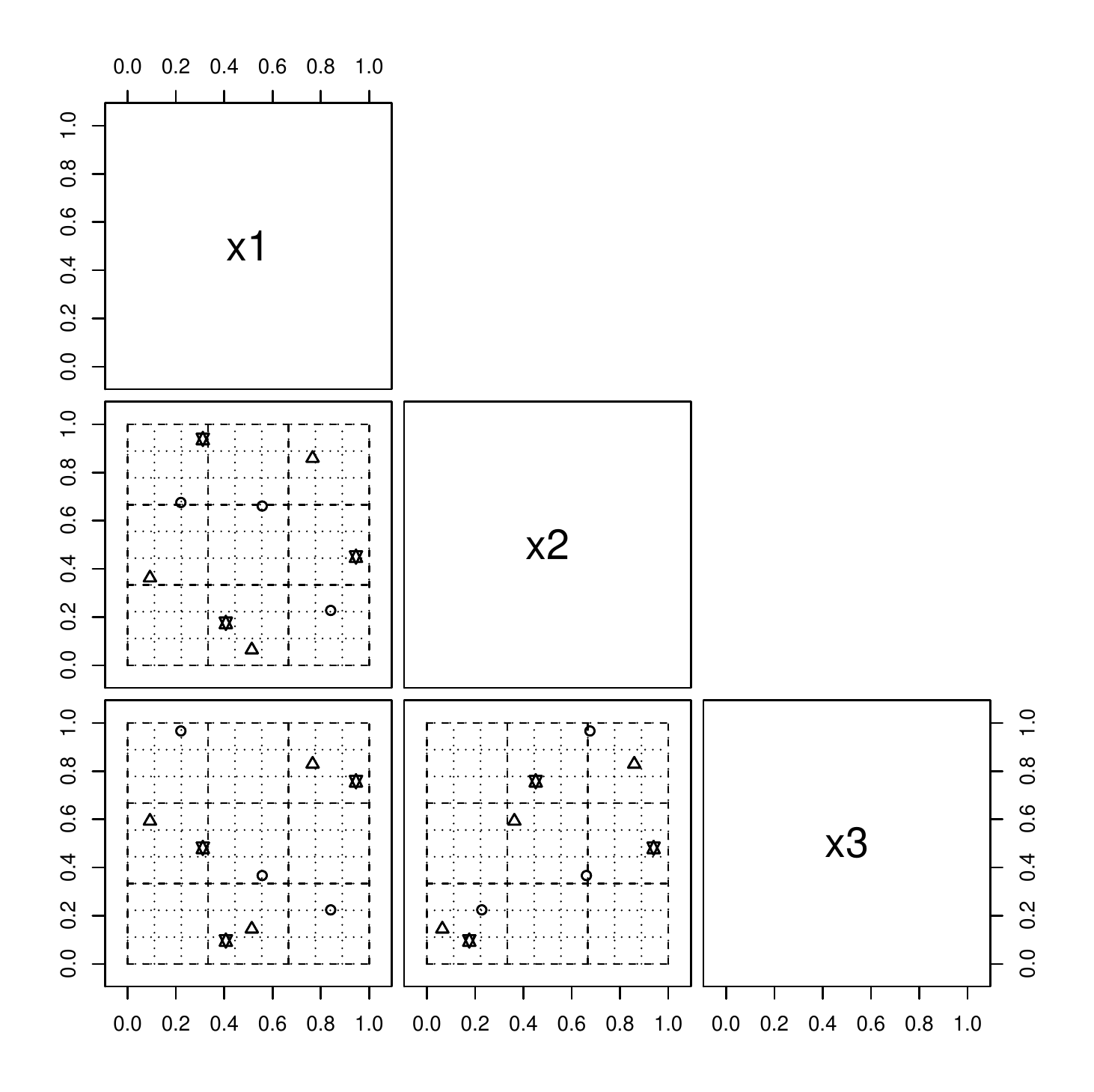}
\caption{Bivariate projections of a sliced Latin hypercube design that
  consists of three $3 \times 3$ ordinary Latin hypercube designs,
  each denoted by a different symbol.}
\label{fig:slhd}
\end{figure}

Figure~\ref{fig:slhd} displays the bivariate projection of the three
$3 \times 3$ ordinary Latin hypercube designs, each denoted by a
different symbol, which are generated under an SLH scheme. For each
design, each of the three equally spaced intervals of $(0, 1]$
  contains exactly one scenario in each dimension. In contrast to
  Figure~\ref{fig:ilhd}, when we combine the three designs together,
  each of the {\em nine} equally spaced intervals of $(0, 1]$ contains
  exactly one scenario in any one dimension.

\subsection{Monotonicity Condition}\label{subsec:samp_slhd}

We derive theoretical results to show $\var(L_{n,t}^{SLH}) \leq
\var(L_{n,t}^{ILH})$ under a monotonicity condition that will be
defined shortly. 


\begin{definition} 
We say that two random variables $Y$ and $Z$ are {\em negatively
  quadrant dependent} if
\[
P(Y \leq y, Z\leq z) \leq P(Y\leq y)P(Z\leq z), \quad \mbox{for all $
  y,z$.}
\]
\end{definition}

\begin{definition}\label{def:block}
Let $B=(b_{ik})$ denote the underlying ordinary Latin hypercube of $D$
in \eqref{eq:A-D} such that $D=(B-\Theta)/n$. Let $v_n(D) =
H(\Delta;B)$ given $B$, where $H(\Delta;B):(0,1]^{n \times m}
\rightarrow \mathbb{R}$ and $\Delta=(\delta_{\ell k})$, with
$\delta_{\ell k}=\theta_{ik}$ such that $b_{ik}=\ell$ for
$\ell=1,2,\dotsc,n$ and $k=1,2,\dotsc,m$. The function $v_n(D)$ is
said to be {\em monotonic} if the following two conditions hold: (a)
for all $B$, $H(\Delta;B)$ is monotonic in each argument of $\Delta$,
and (b) the direction of the monotonicity in each argument of $\Delta$
is consistent across all $B$.
\end{definition}

\begin{example}\label{ex:block}
Consider $D=(\xi_{ik})_{3 \times 2}$. Let
$v_n(D)=\sum_{i=1}^3\sum_{k=1}^2 (\xi_{ik}-1/3)^2$. When
\begin{equation*}
B=
\left[
\begin{array}{cc}
3 & 1 \\
2 & 3\\
1 & 2\\
\end{array}\right], 
\end{equation*}
we have $\delta_{11}=\theta_{31}$, $\delta_{21}=\theta_{21}$,
$\delta_{31}=\theta_{11}$, $\delta_{12}=\theta_{12}$,
$\delta_{22}=\theta_{32}$, and $\delta_{32}=\theta_{22}$ and
\begin{equation*}
H(\Delta;B)= \frac19 \left[ \delta_{11}^2+(1-\delta_{21})^2+(2-\delta_{31})^2+\delta_{12}^2+(1-\delta_{22})^2+(2-\delta_{32})^2 \right],
\end{equation*}
Thus, $H(\Delta;B)$ is increasing in $\delta_{11}$ and $\delta_{12}$
while it is decreasing in the other $\delta_{ik}$s, for $\delta_{ik}
\in (0,1]$. This is true for any underlying ordinary Latin hypercube
  $B$, since $\delta_{11}$ and $\delta_{12}$ are always associated
  with values in $D$ which are between (0,1/3] in this example. By
    definition, $v_n(D)$ is monotonic.  
\end{example}

The monotonicity condition can be checked by directly studying the
function $v_n(D)$, but it can also be shown to be satisfied if the
stochastic program has certain nice properties. Later we will prove
some sufficient conditions for two-stage stochastic linear programs
and give a simple argument to show how some problems from the 
literature have the monotonicity property.

\cite{Qian2012} proves that the function values of any two scenarios in a
sliced Latin hypercube design are negatively quadrant dependent. The
next lemma extends this result, showing the function values of any two
\emph{batches} in a sliced Latin hypercube design are also negatively
quadrant dependent, under the monotonicity assumption on $v_n(D)$.
\begin{lemma}\label{lm:block}
Consider $D_1,D_2,\dotsc,D_t$ generated by Algorithm~\ref{alg:slhd}. If $v_n(D)$ is monotonic, then we have
\begin{equation}
\mathbb{E}[v_n(D_r)v_n(D_s)] \leq \mathbb{E}[v_n(D_r)]
\mathbb{E}[v_n(D_s)] \nonumber,
\end{equation}
for any $r,s=1,2,\dotsc,t$ with $r \neq s$.
\end{lemma}
\proof  Given two ordinary Latin hypercubes $B_{r}$ and
$B_{s}$ in \eqref{eq:B-D}, let $D_{r}=(B_{r}-\Theta_{r})/n$ and
$D_{s}=(B_{s}-\Theta_{s})/n$ denote two slices generated by
\eqref{eq:slhd3}. Since the underlying Latin hypercubes are fixed for
$D_{r}$ and $D_{s}$, the only random parts in the definition are
$\Theta_{r}$ and $\Theta_{s}$. Define
$H(\Delta_{r};B_{r})=v_n(D_{r})=v_n(B_{r}/n-\Theta_{r}/n)$ and
$H(\Delta_{s};B_{s})=v_n(D_{s})=v_n(B_{s}/n-\Theta_{s}/n)$, where $\Delta_{r}$ and $\Delta_{s}$ are $n
\times m$ matrices with the $(k,\ell)$th entry defined as $\delta_{r,
  \ell k}=\theta_{r,ik}$ and $\delta_{s,\ell k}=\theta_{s,ik}$ such
that $B_{r,ik}=B_{s,ik}=\ell$ for $\ell=1,2,\dotsc,n$ and
$k=1,2,\dotsc,m$.  By Proposition~\ref{prop:sigma} (iii) and (iv), we can find $nm$ independent pairs of random variables:
$(\delta_{r},\delta_{s})$ for $\ell=1,2,\dotsc,n$,
$k=1,2,\dotsc,m$. By \cite[Theorem~2]{Lehmann1966}, the monotonicity
assumption of $v_n(D)$, and the proof of \cite[Theorem~1]{Qian2012},
we have
\[
\mathbb{E}\left[H(\Delta_{r};B_{r})H(\Delta_{s};B_{s})\right] \leq
\mathbb{E}\left[H(\Delta_{r};B_{r})\right] \,
\mathbb{E}\left[H(\Delta_{r};B_{r})\right],
\]
which is equivalent to
\[
\mathbb{E}\left[v_n(D_{r})v_n(D_{s})|B_{r},B_{s} \right] \leq
\mathbb{E}\left[v_n(D_{r})|B_{r} \right] \,
\mathbb{E}\left[v_n(D_{r})|B_{s} \right]. 
\]
Taking expectations of both sides gives
\[
  \mathbb{E} \left[v_n(D_{r})v_n(D_{s}) \right] \leq
  \mathbb{E} \left\{ \mathbb{E} \left[v_n(D_{r})|B_{r} \right] \right\}
  \mathbb{E} \left\{ \mathbb{E} \left[v_n(D_{s})|B_{s} \right] \right\}=\mathbb{E} \left[ v_n(D_{r}) \right]
  \mathbb{E} \left[ v_n(D_{s}) \right].
\]
The last equality holds because $B_r$ and $B_s$ are independent, by
Proposition~\ref{prop:sigma} (i). \Halmos \endproof

The next result is an immediate consequence of
Lemma~\ref{lm:block}. It indicates that the variance of the
lower-bound estimator could be reduced by using SLH, when $v_n(D)$ is
monotonic.
\begin{theorem}\label{thm:block}
Consider two lower bound estimators $L_{n,t}^{ILH}$ and $L_{n,t}^{SLH}$
in (\ref{eq:lbe}) obtaind under ILH and SLH, respectively. Suppose
$v_n(D)$ in monotonic, then for any $n$ and $t$, we have that
\[
\var(L_{n,t}^{SLH}) \leq \var(L_{n,t}^{ILH}).
\]
\end{theorem}
%


Even if the monotonicity condition does not hold, we can prove similar
statements about the asymptotic behavior of the variance of ILH and SLH.

Theorem \ref{thm:large_t} gives an asymptotic result that shows the same
conclusion can be drawn as in Theorem \ref{thm:block} even if the
monotonicity condition does not hold. 

\begin{theorem}\label{thm:large_t}
Let $D$ denote an $n \times m$ Latin hypercube design based on a Latin
hypercube $B$ such that $D=(B-\Theta)/n$. Let $H(\Delta;B)=v_n(D)$. Suppose
    $\mathbb{E}\{[v_n(D)]^2\}$ is well defined. As $t \rightarrow
    \infty$ with $n$ fixed, the following are true.
\begin{enumerate}
\item[(i)] $\var(L_{n,t}^{ILH})=t^{-1}\var[v_n(D)]$.
\item[(ii)]
  $\var(L_{n,t}^{SLH})=t^{-1}\var[v_n(D)]-t^{-1}\sum_{\ell=1}^n\sum_{k=1}^m
  \int \{\mathbb{E}[H_{\{\ell k\}}(\delta_{\ell k};B)]\}^2
  d\delta_{\ell k}+o(t^{-1})$, where $H_{\{\ell k\}}(\delta_{\ell
    k};B)$ is the main effect function of $H(\Delta;B)$ with respect
  to $\delta_{\ell k}$.
\item[(iii)]
If $v_n(D)=\sum_{k=1}^m v_{n,\{k\}}(D(:,k))$ is additive, where $v_{n,\{k\}}:(0,1]^n \rightarrow \mathbb{R}$, then $\var(L_{n,t}^{SLH})=o(t^{-1})$.
\end{enumerate}
\end{theorem}

\proof  (i) When $D_1,D_2,\dotsc,D_t$ are sampled independently
under ILH, $\cov[v_n(D_r),v_n(D_s)]=0$ for any $r \neq s$. Thus, we 
have $\var(L_{n,t}^{ILH})=t^{-1}\var[v_n(D)]$. 

(ii) When $D_1,D_2,\dotsc,D_t$ are sampled under SLH, we have
\begin{align}\label{eq:asy_slhd1}
\var(L_{n,t}^{SLH})=&\var \left[ t^{-1}\sum_{i=1}^t v_n(D_i) \right]
\nonumber\\ =&t^{-2}\sum_{i=1}^t \var \left[ v_n(D_i) \right]+t^{-2}\sum_{1\leq
  r<s\leq t} \cov \left[ v_n(D_r)v_n(D_s) \right] \nonumber\\ =&
\var(L_{n,t}^{ILH})+(1-t^{-1}) \cov \left[ v_n(D_1)v_n(D_2) \right],
\end{align}
where the last equality in \eqref{eq:asy_slhd1} holds because batches
are exchangeable. Let $H(\Delta_1;B_1)=v_n(D_1)$ and
$H(\Delta_2;B_2)=v_n(D_2)$. Without loss of generality, assume
$\mathbb{E}[v_n(D)]=0$. We have
\begin{align}
\cov[v_n(D_1)v_n(D_2)]=&\mathbb{E} \left[ v_n(D_1)v_n(D_2) \right] \nonumber\\
=& \mathbb{E}\left\{ \mathbb{E} \left[ v_n(D_1)v_n(D_2)|B_1,B_2 \right] \right\} \nonumber\\
=& \mathbb{E} \left\{ \mathbb{E}\left[ H(\Delta_1;B_1)H(\Delta_2;B_2)\right] \right\} \nonumber\\
=& \mathbb{E} \left[ \sum_{\ell=1}^n\sum_{k=1}^m -t^{-1} \int H_{\{\ell k\}}(\delta_{\ell k};B_1) H_{\{\ell k\}}(\delta_{\ell k};B_2) d\delta_{\ell k}\right]+o(t^{-1}) \nonumber \\
=& -t^{-1}\sum_{\ell=1}^n\sum_{k=1}^m \int \left \{\mathbb{E}  \left[ H_{\{\ell k\}}(\delta_{\ell k};B) \right] \right\}^2 d\delta_{\ell k}+o(t^{-1}), \label{eq:asy_slhd3}
\end{align}
where the second-last equality is from \cite[Lemma~1]{Qian:2009}. The
result follows by substituting \eqref{eq:asy_slhd3} into
\eqref{eq:asy_slhd1}. The functional ANOVA decomposition is properly
defined because the quantities $\delta_{\ell k}$ are independent,
according to Proposition~\ref{prop:sigma} (iii).

(iii) Assuming again that $\mathbb{E}[v_n(D)]=0$, we have
\begin{align}
\var[v_n(D)]=& \mathbb{E} \left\{ \var \left[v_n(D)|B \right] \right\} + \var
\left\{ \mathbb{E} \left[ v_n(D)|B \right] \right\} \nonumber\\ 
=&
\mathbb{E} \left\{ \var \left[H(\Delta;B) \right] \right\} \nonumber\\ 
=& \mathbb{E} \left\{\sum_{\ell=1}^n\sum_{k=1}^m
\int \left[ H_{\{\ell k\}}(\delta_{\ell k};B) \right]^2 d\delta_{\ell k}\right\}
\nonumber\\ 
=& \sum_{\ell=1}^n\sum_{k=1}^m \int \mathbb{E} \left\{ \left[H_{\{\ell
    k\}}(\delta_{\ell k};B) \right]^2 \right\} d\delta_{\ell k} \nonumber\\ 
=&
\sum_{\ell=1}^n\sum_{k=1}^m \int \var \left[ H_{\{\ell k\}}(\delta_{\ell k};B) \right]+
\left\{\mathbb{E} \left[ H_{\{\ell k\}}(\delta_{\ell k};B) \right ] \right\}^2 d\delta_{\ell
  k}\nonumber\\ =& \sum_{\ell=1}^n\sum_{k=1}^m \int \left\{ \mathbb{E} \left[ H_{\{\ell
    k\}}(\delta_{\ell k};B) \right] \right\}^2 d\delta_{\ell
  k}. \label{eq:asy_slhd4}
\end{align}
The second equality
holds because $\mathbb{E}[v_n(D)|B]$ is the same regardless of the
underlying ordinary Latin hypercube $B$ when $v_n(D)$ is additive. The
third equation holds due to the functional ANOVA decomposition on
$H(\Delta;B)$. We complete the proof by combining \eqref{eq:asy_slhd4}
with the result in (ii). \Halmos
\endproof

\subsection{Two-Stage Linear Program}\label{subsec:2stage}

We now discuss the theoretical properties of SLH for two-stage
stochastic linear programs. Consider problems of the form
\begin{equation}\label{eq:two_stage}
\min_{\substack{x \in X}} c^T x + \mathbb{E}[Q(x,\xi)],
\end{equation}
where $X$ is a polyhedron and
\begin{equation}\label{eq:prime}
Q(x,\xi)=\text{inf} \, \left\{ q^T y \, : \, W y \leq h-Tx, \; y \geq 0 \right\},
\end{equation}
and $\xi:=(h,q,T)$.  The problem has {\em fixed recourse} since the
recourse matrix $W$ does not depend on the random variable $\xi$.
Defining $G(x,\xi)$ to be the function $c^T x+Q(x,\xi)$, we see that
\eqref{eq:two_stage} is a special case of \eqref{eq:sp}. Let
$x=(x^1,x^2,\dotsc,x^p)$ and $T=(T_{kj})$. By \eqref{eq:prime},
$Q(x,\xi)$ is a decreasing function of any entry in $h$, for any $x
\in X$. Furthermore, $Q(x,\xi)$ is a decreasing function of any entry
$T_{kj}$ in $T$ if $x^j$ is nonpositive, and an increasing function of
any entry $T_{kj}$ in $T$ if $x^j$ is nonnegative.

By LP duality,  we have
\[
Q(x,\xi)=\text{sup} \, \left\{ u^T (h-Tx) \,: \, W^T u \leq q, \; u \leq 0 \right\},
\]
and hence $Q(x,\xi)$ is an increasing function of any entry in $q$ for
any $x \in X$. We conclude that $G(\cdot,\xi)$ is monotonic in each
component of $\xi$ if the recourse matrix $W$ is fixed.

\begin{lemma}\label{lm:two_mono}
Let $v_n(D)$ in \eqref{eq:spn} denote the approximated optimal value
of the two-stage stochastic program in \eqref{eq:two_stage} with fixed
recourse. Then $v_n(D)$ is monotonic if
\emph{(i)} $T$ is fixed or \emph{(ii)} for every $j=1,2,\dotsc,p$,
$x^j$ is always nonnegative or nonpositive, given any $D$.
\end{lemma}
\proof  For any ordinary Latin hypercube design $D$, let $x_n$
be the arg min of the approximation problem, that is,
\[
v_n(D)=\min_{x \in X} \, \frac{1}{n}\sum_{i=1}^n
G(x,\xi_i)=\frac{1}{n}\sum_{i=1}^n G(x_n,\xi_i),
\]
where $\xi_i$ is the $i$th scenario of $D$. Without loss of generality,
increase only $\xi_{1k}$ to obtain a new design $D^*$ with scenarios
$\xi_1^*,\xi_2,\dotsc,\xi_n$. Let $x_n^*$ be the arg min of the
approximation problem with design $D^*$, that is,
\[
v_n(D^*)=\min_{x \in X} \, \frac{1}{n} \left[ G(x,\xi_1^*)+\sum_{i=2}^n
    G(x,\xi_i) \right]=\frac{1}{n} \left[ G(x_n^*,\xi_1^*)+\sum_{i=2}^n
    G(x_n^*,\xi_i) \right].
\]
If either (i) or (ii) is satisfied, the value of $x$ does not affect
whether $G(\cdot,\xi_1)$ is increasing or decreasing in
$\xi_{1k}$. Supposing that $G(\cdot,\xi_1)$ is increasing in
$\xi_{1k}$, we have
\begin{equation}
v_n(D^*)=\frac{1}{n} \left[ G(x_n^*,\xi_1^*)+\sum_{i=2}^n G(x_n^*,\xi_i) \right]
\geq \frac{1}{n}\sum_{i=1}^n G(x_n^*,\xi_i) \geq v_n(D), \nonumber
\end{equation}
which implies that $v_n(D)$ is increasing in $\xi_{1k}$. Similarly, if
$f(\cdot,\xi_1)$ is decreasing in $\xi_{1k}$, then
\begin{equation}
v_n(D^*) \leq \frac{1}{n} \left[ G(x_n,\xi_1^*)+\sum_{i=2}^n G(x_n,\xi_i) \right] \leq \frac{1}{n}\sum_{i=1}^n G(x_n,\xi_i) =v_n(D), \nonumber
\end{equation}
which implies $v_n(D)$ is decreasing in $\xi_{1k}$. \Halmos
\endproof

\begin{example}\label{ex:news}
Consider the newsvendor problem from \cite{Freimer2012}, which can be
expressed as a two-stage stochastic program. In the first stage,
choose an order quantity $x$. After demand $\xi$ has been realized, we
decide how much of the available stock $y$ to sell. Assume that demand
is uniformly distributed on the interval $(0, 1]$, and there is a
  shortage cost $\alpha \in (0,1)$ and an overage cost $1-\alpha$. The
  second stage problem is
\[
P: \quad\quad Q(x,\xi)=\min_y \, \left[ (1-\alpha)(x-y)+\alpha(\xi-y) \, | \, y \leq x, \; y\leq \xi \right].
\]
Since the first-stage cost is zero, the two-stage stochastic program is
\[
MP: \quad\quad \min_x  \, \mathbb{E} \left\{ \min_y \, \left[ (1-\alpha)(x-y)+\alpha(\xi-y) \, | \, y \leq x, \; y\leq \xi \right] \right\}.
\]
The optimal value is $v^*={\alpha(1-\alpha)}/{2}$ with solution
$x^*=\alpha$.

Based on a sample of $n$ demands $\xi_1,\xi_2,\dotsc,\xi_n$, the approximated
optimal value is given by
\[
v_n(D)= \min_x \, \frac{1}{n} \sum_{i=1}^n \left[ (1-\alpha)(x-\xi_i)^++\alpha(\xi_i-x)^+ \right],
\]
where $D=(\xi_1,\xi_2,\dotsc,\xi_n)^T$. The optimal solution
$x_n^*$ is the $\left\lceil \alpha n \right\rceil$th order statistic
of $\{\xi_1,\xi_2,\dotsc,\xi_n\}$.

\begin{table}[htbp]
\centering
\caption{Means and standard errors (in parentheses) with 1000 replicates}
\begin{tabular}{ccccc}
\hline
$n$ & Scheme & $t=5$ & $t=10$ & $t=20$\\
\hline
\multirow{2}{*}{2} & ILH & 0.1003 (1.83E-2) & 0.1002 (1.31E-2) & 0.1000 (9.23E-3)\\
& SLH & 0.0999 (3.71E-3) & 0.1000 (1.31E-3) & 0.1000 (4.49E-4)\\
\hline
\multirow{2}{*}{20} & ILH & 0.1201 (6.99E-4) & 0.1200 (5.01E-4) & 0.1200 (3.70E-4) \\
& SLH & 0.1200 (1.43E-4) & 0.1200 (4.87E-5) & 0.1200 (1.72E-5)\\
\hline
\multirow{2}{*}{200} & ILH & 0.1200 (2.18E-5) & 0.1200 (1.60E-5) & 0.1200 (1.14E-5)\\
& SLH & 0.1200 (4.40E-6) & 0.1200 (1.59E-6) & 0.1200 (5.60E-7)\\
\hline
\end{tabular}
\label{tb:news}
\end{table}

Let $\alpha=.4$, for which $v^*=0.12$. Table~\ref{tb:news} gives means
and standard errors for the estimators of $L_{n,t}=\mathbb{E}[v_n(D)]$
for several values of $n$ and $t$.  This table shows that SLH reduces
the variance of $L_{n,t}$ significantly when compared with
ILH. Analytically, we have
\[
v_n(D) = H(\Delta;B)=n^{-2}\sum_{i=1}^{r^*-1}
(1-\alpha)(r^*-i+\delta_{r^*}-\delta_i)+n^{-2}\sum_{i=r^*+1}^n
\alpha(i-r^*+\delta_i-\delta_{r^*}),
\]
where $r^*=\left\lceil \alpha n \right\rceil$ and $B$ is an arbitrary
underlying Latin hypercube for $D$. We notice that for any $B$,
$H(\Delta;B)$ is decreasing in $\delta_1,\ldots,\delta_{r^*-1}$,
increasing in $\delta_{r^*+1},\ldots,\delta_{n}$ and monotonic in
$\delta_{r^*}$ (the direction depends on $\alpha$). Thus, $v_n(D)$ is
monotonic, which can alternatively be checked by applying Lemma
\ref{lm:two_mono}. By
Theorem~\ref{thm:block}, we should have $\var(L_{n,t}^{SLH}) \leq
\var(L_{n,t}^{ILH})$.

We notice that $\var(L_{n,t}^{ILH})=O(t^{-1})$ while
$\var(L_{n,t}^{SLH})=o(t^{-1})$, a fact that can be explained by
Theorem~\ref{thm:large_t} (iii), since the newsvendor problem only has
one random variable and $v_n(D)$ is additive.
\end{example}

\subsection{Discrete Random Variables}\label{subsec:link}

Theorems~\ref{thm:block} and \ref{thm:large_t} confirm the
effectiveness of sliced Latin hypercube designs in reducing the
variance of lower-bound estimates in SAA.  However, the assumptions in those
theorems limit their applicability to fairly specialized problems.
Theorem~\ref{thm:block} does not apply to two-stage problem in
\eqref{eq:two_stage} with random recourse. Theorem~\ref{thm:large_t} does not apply when $n \rightarrow \infty$,
which is a more practical assumption than $t \rightarrow \infty$. In
this section, we consider the case in which $\xi^1,\xi^2,\dotsc,\xi^m$
are discrete random variables, as discussed by \cite{Mello2008}. In
fact, we assume $\xi^1,\xi^2,\dotsc,\xi^m$ to be {\em independent}
discrete random variables. We plan to show that estimating the lower
bound of $v^*$ is almost equivalent to a numerical integration
problem. Several existing results regarding numerical integration
provide us with tools for studying effects of different sampling
schemes for lower bound estimation more generally.

\begin{assumption}\label{ass:consist}
For each $x \in X$, $\hat{g}_n(x) \rightarrow g(x)$ with probability
one (denoted ``w.p. 1'').
\end{assumption}

\begin{assumption}\label{ass:linear}
The feasible set $X$ is compact and polyhedral; the
function $G(\cdot,\xi)$ is convex piecewise linear for every value of
$\xi$; and the distribution of $\xi$ has finite support.
\end{assumption}

Assumption~\ref{ass:consist} holds under Latin hypercube sampling if
$\mathbb{E} \left\{[G(x,\xi)]^2 \right\}<\infty$ for every $x \in X$,
by \cite[Theorem~3]{Loh1996}. Assumption~\ref{ass:linear} holds in
practice for stochastic linear programs in which the random vector is
discretized. Let $S^*$ denote the set of optimal solutions of (\ref{eq:sp}). Based on both assumptions, \cite{Mello2008} shows the
consistency of the approximated optimal solution in \eqref{eq:spn}.

\begin{proposition} {\rm \cite[Proposition~2.5]{Mello2008}.}
\label{prop:mello}
Suppose that Assumptions~\ref{ass:consist} and \ref{ass:linear}
hold. Then $x_n^* \in S^*$ w.p. 1, for $n$ sufficiently large.
\end{proposition} 

Now let $F_k$ denote the cumulative distribution function of $\xi^k$
for $k=1,2,\dotsc,m$. Define the inverse of $F_k$ as
$F^{-1}_k(z):=\inf \{y \in \Xi_k \, : \, F_k(y) \geq z\}$, where
$\Xi_k$ is the support of $F_k$ and $z \in 0,1]$. We
  can express $\xi=(\xi^1,\xi^2,\dotsc,\xi^m)$ as
  $\Psi(\tilde\xi)=(F_1^{-1}(\tilde\xi^1),F_2^{-1}(\tilde\xi^2),\dotsc,F_m^{-1}(\tilde\xi^m)))$,
  where $\tilde\xi$ is a random vector uniformly distributed on
  $(0,1]^m$. Define $\tilde{G}(x,\tilde{\xi})$ on $(0,1]^m$ so that
      $\tilde{G}(x,\cdot)=G(x,\Psi(\cdot))$. Results obtained in
      previous sections would still apply with respect to $\tilde{G}$
      and $\tilde\xi$.

The following proposition connects two-stage stochastic linear program
in \eqref{eq:two_stage} and numerical integration.

\begin{proposition}\label{prop:equal}
Consider problem \eqref{eq:two_stage}, and suppose that $\mathbb{E}
\left\{ [\tilde{G}(x,\tilde\xi)]^2 \right\}<\infty$ for every $x \in
X$, and that $S^*=\{x^*\}$ is a singleton. Then for $n$ sufficiently
large, we have 
\[
L_{n,t}=(nt)^{-1} \sum_{r=1}^t
\sum_{i=1}^n\tilde{G}(x^*,\tilde\xi_{r,i}) \quad w.p.1,
\]
where $\tilde\xi_{r,i}$ is the $i$th scenario in $D_r$.
\end{proposition}
\proof  Let $x^*_n(D_r)$ denote an optimal solution to the SAA
problem with scenarios given by $D_r$ for $r=1,2,\dotsc,t$, we have $L_{n,t}=t^{-1}\sum_{r=1}^t
v_n(D_r)=(nt)^{-1}\sum_{r=1}^t\sum_{i=1}^n
\tilde{G}(x^*_n(D_r),\tilde\xi_{r,i})$. Because $x^*_n(D_r)=x^*$ w.p.1 for $n$
sufficiently large by Proposition~\ref{prop:mello} and $\tilde{G}(\cdot,\tilde{\xi})$ is continuous, the result follows according to the Continuous Convergence Theorem \citep{Mann1943}.  \Halmos
\endproof

Proposition~\ref{prop:equal} indicates that $L_{n,t}$ becomes an
integral estimator of $\tilde{G}(x^*,\xi_{r,i})$ and that $L_{n,t}$ is
directly estimating the true optimal value $v^*$ of \eqref{eq:sp}, for
$n$ large enough. Results about the variance formula for
$L_{n,t}^{ILH}$ and $L_{n,t}^{SLH}$ are given in the next result.
\begin{theorem}\label{thm:var_slhd}
Consider problem \eqref{eq:two_stage}. With the same conditions in
Proposition~\ref{prop:equal}, we have that the following results hold
as $n,t \rightarrow \infty$:
\begin{enumerate}
\item [\emph{(i)}] $\var(L_{n,t}^{ILH})=(nt)^{-1}\var
  \left[\tilde{G}(x^*,\tilde\xi) \right]-(nt)^{-1}\sum_{k=1}^m\var
  \left[ \tilde{G}_{\{k\}}(x^*,\tilde\xi^k) \right]+o(n^{-1}t^{-1});$
\item [\emph{(ii)}] $\var(L_{n,t}^{SLH})=(nt)^{-1}\var
  \left[\tilde{G}(x^*,\tilde\xi) \right]-(nt)^{-1}\sum_{k=1}^m\var
  \left[\tilde{G}_{\{k\}}(x^*,\tilde\xi^k) \right]+o(n^{-1}t^{-1});$
\item [\emph{(iii)}] If $\tilde{G}(x^*,\tilde\xi)=\sum_{k=1}^m \tilde{G}_{\{k\}}(x^*,\tilde\xi^k)$ is additive, then
  $\var(L_{n,t}^{SLH}) \leq \var(L_{n,t}^{ILH})$. Furthermore, we have
\begin{equation*}
\var(L_{n,t}^{ILH})=O(n^{-2}t^{-1}) \;\; \mbox{\rm and} \;\;
\var(L_{n,t}^{SLH})=O(n^{-2}t^{-2}).
\end{equation*}
\end{enumerate}
\end{theorem}  

\proof  (i) Since $D_1,D_2,\dotsc,D_t$ are independent
ordinary Latin hypercube designs sampled under ILH, we immediately
have that $\var(L_{n,t}^{ILH})=t^{-1}\var \left[v_n(D_1) \right]$, by
exchangeability of batches. By \cite[Corollary~1]{Stein1987}, we have
\[
\var \left[v_n(D_1) \right]=n^{-1}\var \left[ \tilde{G}(x^*,\tilde\xi)
  \right]-n^{-1}\sum_{k=1}^m\var \left[
  \tilde{G}_{\{k\}}(x^*,\tilde\xi^k) \right]+o(n^{-1}),
\]
and the result follows.

(ii) The result holds as a consequence of \cite[Theorem~2]{Qian2012}.

(iii) Since any batch sampled under SLH is statistically equivalent to
an ordinary Latin hypercube design, by Proposition~\ref{prop:sigma}
(v), the variances $\var \left[ v_n(D_i) \right]$ are the same as
those under ILH. Due to exchangeability of batches and scenarios within the
same batch, it suffices to show that under SLH, we have $\cov
\left[\tilde{G}(x^*,\tilde\xi_{1,1}),\tilde{G}(x^*,\tilde\xi_{2,1})
  \right] \leq 0$, where $\tilde\xi_{r,i}$ denote the $i$th scenario in
$D_r$.  For $0<z_1,z_2\leq 1$ and an integer $p$, define
\[
\alpha_p(z_1,z_2)=
\begin{cases} 
1, &\lceil pz_1 \rceil=\lceil pz_1 \rceil\\ 0, &\mbox{o.w.}
\end{cases}
\]
By \cite[Lemma~1 (iii)]{Qian2012}, the joint probability density
function of $\tilde\xi_{1,1}$ and $\tilde\xi_{2,1}$ is
\begin{equation}
\left(\frac{t}{t-1}\right)^m \prod_{k=1}^m \left[ 1-t^{-1}+t^{-1} \alpha_n(\tilde\xi_{1,1k},\tilde\xi_{2,1k})-\alpha_{nt}(\tilde\xi_{1,1k},\tilde\xi_{2,1k}) \right].
\end{equation} 
Let $I(i,n)$ denote the interval $((i-1)/n,i/n]$. Without loss of
  generality, assume that $\mathbb{E}[\tilde{G}(x^*,\tilde\xi)]=0$,
  where $\tilde{G}(x^*,\tilde\xi)=\sum_{k=1}^m
  \tilde{G}_{\{k\}}(x^*,\tilde\xi^k)$.  Following the proof of
  \cite[Theorem~1]{Stein1987}, we can express $\cov
  \left[\tilde{G}(x^*,\tilde\xi_{1,1}),\tilde{G}(x^*,\tilde\xi_{2,1})
    \right]$ as
\begin{align}
\cov \left[
  \tilde{G}(x^*,\tilde\xi_{1,1}),\tilde{G}(x^*,\tilde\xi_{2,1})
  \right] =&\frac{t}{t-1} \sum_{k=1}^m \left(\frac{1}{t}\int
\tilde{G}_{\{k\}}(x^*,\tilde\xi_{1,1k})\tilde{G}_{\{k\}}(x^*,\tilde\xi_{2,1k})\alpha_{n}(\tilde\xi_{1,1k},\tilde\xi_{2,1k})d\tilde\xi_{1,1k}d\tilde\xi_{2,1k}
\right. \nonumber \\ 
& \quad \left. -\int\tilde{G}_{\{k\}}(x^*,\tilde\xi_{1,1k})\tilde{G}_{\{k\}}(x^*,\tilde\xi_{2,1k})\alpha_{nt}(\tilde\xi_{1,1k},\tilde\xi_{2,1k})d\tilde\xi_{1,1k}d\tilde\xi_{2,1k}\right)
\nonumber\\ =&\frac{t}{t-1} \sum_{k=1}^m \left[\frac{1}{t}\sum_{i=1}^n
  \left(\int_{I(i,n)}\tilde{G}_{\{k\}}(x^*,\tilde\xi^k)d\tilde\xi^k\right)^2\right. \nonumber\\ & \quad \left. -\sum_{i=1}^{nt}
  \left(\int_{I(i,nt)}\tilde{G}_{\{k\}}(x^*,\tilde\xi^k)d\tilde\xi^k\right)^2
  \right] \label{eq:slh1d}
\end{align}
Notice that $I(i,n)=\cup_{j=(i-1)t+1}^{it} I(j,nt)$ for any
$i=1,2,\dotsc,n$. By Jensen's inequality, for each $k=1,2,\dotsc,m$
and $i=1,2,\dotsc,n$, we have
\begin{equation*}
\frac{1}{t}\left(\int_{I(i,n)} \tilde{G}_{\{k\}}(x^*,\tilde\xi^k)d\tilde\xi^k\right)^2 \leq \sum_{j=(i-1)t+1}^{it}\left(\int_{I(j,nt)} \tilde{G}_{\{k\}}(x^*,\tilde\xi^k)d\tilde\xi^k\right)^2.
\end{equation*}
The proof of the first claim is completed by substituting this
inequality into \eqref{eq:slh1d}.

To prove the second claim, we define
\begin{equation*}
\zeta_i^n(\tilde\xi^k)=\tilde{G}_{\{k\}}(x^*,\tilde\xi^k)-\int_{I(i,n)} \tilde{G}_{\{k\}}(x^*,\tilde\xi^k)d\tilde\xi^k.
\end{equation*}
Again, by following the proof of \cite[Theorem~1]{Stein1987}, we have
\begin{equation*}
\var[v_n(D_1)]=\frac{1}{n}\sum_{k=1}^m \sum_{i=1}^n \left[
  \zeta_i^n(\tilde\xi^k) \right]^2.
\end{equation*}
Because each $\Xi_k$ is finite, the number of non-zero values of
$\zeta_i^n(\tilde\xi^k)$ is also finite. Hence,
$\var(L_{n,t}^{ILH})=t^{-1}\var[v_n(D_1)]=O(n^{-2}t^{-1})$. Under SLH,
using the same arguments as \eqref{eq:slh1d}, we have
\[
\cov \left[ \tilde{G}(x^*,\tilde\xi_{1,1}),\tilde{G}(x^*,\tilde\xi_{2,1}) \right]
=\frac{t}{t-1}
\sum_{k=1}^m \left[\frac{1}{nt}\sum_{i=1}^{nt}
  (\zeta_i^{nt}(\tilde\xi^k))^2-\frac{1}{nt}\sum_{i=1}^{n}
  (\zeta_i^{n}(\tilde\xi^k))^2\right], 
\]
and
\begin{align*}
\var(L_{n,t}^{SLH})=&t^{-1}\var \left[v_n(D_1) \right]+
\frac{t-1}{t}\cov \left[
  \tilde{G}(x^*,\tilde\xi_{1,1}),\tilde{G}(x^*,\tilde\xi_{2,1})
  \right] \\ 
=&\frac{1}{nt}\sum_{k=1}^m \sum_{i=1}^n
(\zeta_i^n(\tilde\xi^k))^2+\sum_{k=1}^m
\left[\frac{1}{nt}\sum_{i=1}^{nt}
  (\zeta_i^{nt}(\tilde\xi^k))^2-\frac{1}{nt}\sum_{i=1}^{n}
  (\zeta_i^{n}(\tilde\xi^k))^2\right] \\ 
=&\frac{1}{nt}\sum_{k=1}^m\sum_{i=1}^{nt}
\left(\zeta_i^{nt}(\tilde\xi^k) \right)^2  \\ 
=&O(n^{-2}t^{-2}).\nonumber
\end{align*}
\Halmos
\endproof

Let $\nu_k$ denote the number of all possible distinct values for
$\xi^k$ in $\Xi_k$, and denote by $p_{(i)k}$ the probability of $\xi$
taking its $i$th smallest possible value. To sample $\xi^k$, we will
first take $\tilde\xi^1,\tilde\xi^2,\dotsc,\tilde\xi^m$ in $n$ equally
spaced subintervals in $(0,1]$, respectively. If any $\tilde\xi^k$
  taken from the same subinterval always leads to the same value for
  $\xi^k=F^{-1}_k(\tilde\xi^k)$, then $v_n(\vD)$ will entirely depend
  on the underlying ordinary Latin hypercube $B$. The following
  proposition gives sufficient conditions for equal performance
  between ILH and SLH.
\begin{proposition}\label{prop:pequal}
If $np_{(i)k}$ is an integer for all $i=1,2,\dotsc,\nu_k$ and
$k=1,2,\dotsc,m$, then $\var(L_{n,t}^{SLH})= \var(L_{n,t}^{ILH})$.
\end{proposition} 
\proof  Divide $(0,1]$ into the $n$ equal subintervals as
  $(0,1/n],(2/n,3/n],\dotsc,((n-1)/n,1]$, and let $I(i,n) =
        ((i-1)/n,i/n]$, as before. Assume $p_{(0)k}=0$. For any
          $k=1,2,\dotsc,m$, when $np_{(1)k}$ is an integer, we have
          $\xi^k=F^{-1}_k(\tilde\xi^k)$ equal to the smallest possible
          value in $\Xi_k$, provided that $\tilde\xi^k$ is drawn from
          $I(1,n),I(2,n),\dotsc,I(np_{(1)k},n)$. Similarly, when
          $np_{(2)k}$ is an integer, we have that
          $\xi^k=F^{-1}_k(\tilde\xi^k)$ equal to the second smallest
          value in $\Xi_k$, provided that $\tilde\xi^k$ is taken from
          $I(np_{(1)k}+1,n),I(np_{(1)k}+2,n),\dotsc,I(np_{(2)k},n)$,
          and so on. As a result, $\xi_1,\xi_2,\dotsc,\xi_n$ is
          determined only by $B$. That is, given $b_{ik}$,
          $\tilde\xi_{ik}$ is taken from $I(b_{ik},n)$ and
          $\xi_{ik}=F_k^{-1}(\tilde\xi_{ik})$ is the $\ell$th smallest
          value in $\Xi_k$, where $\ell$ is an integer and
          $np_{(\ell-1)k}+1 \leq b_{ik} \leq n_{(\ell)k}$. Under SLH,
          $v_n(D_{r})$ and $v_n(D_{s})$ would be independent for any
          $r \neq s$, because they depend on $B_{r}$ and $B_{s}$,
          which are independent according to
          Proposition~\ref{prop:sigma} (i).  \Halmos \endproof

Theorem~\ref{thm:var_slhd} (i) and (ii) indicate that ILH and SLH are
equally effective, in general, for estimating $v^*$ when $n,t
\rightarrow \infty$. Proposition~\ref{prop:pequal} gives a specific
case in which SLH is exactly the same as ILH. Another type of sliced
Latin hypercube design is introduced in \S~\ref{sec:oslhd}, which
possesses similar (if not the same) within-batch structure as ILH and
SLH, but much stronger between-batch negative dependence than SLH to
further reduce the variance $\var(L_{n,t})$ of the lower-bound
estimator.

\section{Sliced Orthogonal Array Based Latin Hypercube Sampling}\label{sec:oslhd}

Section~\ref{sec:slhd} indicates that SLH may yield significant
improvement in estimating $L_{n,t}$ when the monotonicity condition in is satisfied or when
$\tilde{G}(x^*,\tilde\xi^k)$ is an additive function in $\tilde\xi^k$,
under Assumptions~\ref{ass:consist} and \ref{ass:linear}.  Both the
monotonicity and additivity conditions emphasize {\em individual}
random variables. As a consequence, the results in
Theorems~\ref{thm:block} and \ref{thm:var_slhd} are intuitive, because
combining all batches under SLH (rather than ILH) gives a design with
better stratification in each dimension; see Figures~\ref{fig:ilhd}
and \ref{fig:slhd}. On the other hand, the combined design under SLH
does not possess better stratification when we consider {\em groups}
of variables. Theorem~\ref{thm:var_slhd} (i) and (ii) suggest that we
would need a better sampling scheme than SLH if neither the
monotonicity nor additivity conditions are satisfied.

We now introduce another sampling scheme with the following
properties.  First, the design for each SAA subproblem in this scheme
is an ordinary Latin hypercube, as for ILH and SLH. Fixing this
property between our different sampling schemes allows us to better
study and understand the benefits of improving the between-batch
stratification. Second, when we choose the number of batches $t$
sufficiently large, the increased size of the combined design matrix
achieves better stratification not just for each variable but also for
every pair of variables.  With the additional stratification, the
$\var(L_{n,t})$ can be further reduced, under the assumptions in
Proposition~\ref{prop:equal}.

\subsection{Orthogonal Arrays}\label{subsec:oslhd_oa}

Orthogonal arrays originate from the pioneering work of
\cite{Rao1946,Rao1947,Rao1949}.  \cite{Patterson1954} introduced
lattice sampling based on randomized orthogonal arrays, which is found
to be suitable for computer experiments and other related fields
\citep{Owen1992}. \cite{Tang1993} and \cite{Owen1994} independently
studied the Monte Carlo variance of means over orthogonal-array-based
Latin hypercube designs and randomized orthogonal arrays,
respectively. 

We introduce some properties and notation for orthogonal arrays that
pertain to Latin hypercube designs. Let $P$ denote a set of $s$
symbols or levels. To be consistent with the notation for a Latin
hypercube $A$ in \S~\ref{subsec:lhd}, we denote levels by
$1,2,\dotsc,s$. The following formal definition of orthogonal arrays
is due to \cite{Hedayat1999}.
\begin{definition}\label{def:oa}
An $N \times m$ array $C=(c_{ik})$ with entries from $P$ is said to be
an orthogonal array with $s$ levels, strength $\tau$ and index
$\lambda$ (for some $\tau$ satisfying $0 \leq \tau \leq m$) if every $N
\times \tau$ subarray of $C$ contains each $\tau-$tuple based on $P$
exactly $\lambda$ times as a row.
\end{definition}

We denote an orthogonal array as OA$(N,m,s,\tau)$, where $N$, $m$,
$s$, $\tau$, and $\lambda$ are all integers, with $N=\lambda
s^\tau$. An ordinary Latin hypercube is an orthogonal array with
$\tau=1$. Furthermore, combined hypercubes under ILH and SLH are
special cases of orthogonal arrays with $\tau=1$. We summarize these
observations in the following proposition.
\begin{proposition}\label{prop:oa_lh}
Let $D_1,D_2,\dotsc,D_t$ denote $t$ Latin hypercube designs of
dimension $n \times m$, which will be used to solve $t$ SAA
problems. Let $D=(\xi_{ik})$ denote the $N \times m$ design by
stacking $D_1,D_2,\dotsc,D_t$ row by row, such that $N=nt$. Let
$A=(a_{ik})$ denote an $N \times m$ array such that $a_{ik}=\lceil n
\xi_{ik} \rceil$. Let $B=(b_{ik})$ denote an $N \times m$ array such
that $b_{ik}=\lceil N \xi_{ik} \rceil$. Then $A$ is an OA$(N,m,n,1)$
under ILH and SLH. Furthermore, $B$ is an OA$(N,m,N,1)$ under SLH.
\end{proposition}

Proposition~\ref{prop:oa_lh} reveals the fact that $L_{n,t}$ is
constructed based on an OA$(N,m,n,1)$ and an OA$(N,m,N,1)$ under ILH
and SLH, respectively.  Theorems~\ref{thm:block}, \ref{thm:large_t},
and \ref{thm:var_slhd} (iii) all indicate that OA$(N,m,N,1)$ is
superior to OA$(N,m,n,1)$ in estimating $L_{n,t}$. In other words, if
two orthogonal arrays have the same values of $N$ and $m$, and have
$\tau=1$, the one with a larger value of $s$ and a smaller value of
$\lambda$ would be more desirable. In the remainder of this section,
we use $\tau=2$, but the same result might still apply when generalized to
larger values of $\tau$.

There exist many methods for constructing orthogonal arrays with
$\tau=2$, most of which use Galois fields and finite geometry. We
summarize some popular constructions and their restrictions for $N$,
$m$, $s$, and $\lambda$ in Table~\ref{tb:oa_con}.

\begin{table}[htbp]
\centering
\caption{Methods of constructing orthogonal arrays with strength two}
\begin{tabular}{lcc}
\hline
Method & Orthogonal Array & Restrictions \\
\hline
\cite{Bush1952} & OA$(s^2,m,s,2)$ & $m \leq s+1$\\
& & $s$ is a prime power \\
\hline
\cite{BoseBush1952} & OA$(\lambda s^2,m,s,2)$ & $m \leq \lambda s+1$\\
& & ($s$ and $\lambda$ are powers of the same prime) \\
\hline
\cite{Addelman1961} & OA$(2s^2,m,s,2)$ & $m \leq 2s+1$\\
& & ($s$ is an odd prime power) \\
\hline
\cite{Addelman1961} & OA$(2s^3,m,s,2)$ & $m \leq 2s^2+2s+1$\\
& & ($s$ is an odd prime power, 2 or 4) \\
\hline
\end{tabular}
\label{tb:oa_con}
\end{table}

Table~\ref{tb:two_oa} presents two strength-two orthogonal
arrays. Both have sixteen scenarios and five columns, the left one having
four levels while the right one has just two levels. Intuitively, the
left one seems preferable, as it includes all 16 possible level
combinations in any two columns. \cite{Owen1994} defines the concept
of coincidence defect, which can be used to more formally justify the
superiority of the left array which has more levels. An orthogonal
array $A$ with strength $\tau$ has \emph{coincidence defect} if there
exist two rows of $A$ that agree in $\tau+1$ columns. The left array
in Table~\ref{tb:two_oa} does not have coincidence defect because no
two rows of $A$ agree in more than a single column. The right one
contains coincidence defects; for example, the second and the third
rows agree in columns 2, 3, and 4.

\begin{table}[htbp]
\setlength{\tabcolsep}{8pt}
\centering
\caption{Two Orthogonal Arrays}
\begin{tabular}{cccccc}
\hline
\multicolumn{6}{c}{An OA$(16,5,4,1)$} \\
\hline
Scenarios\# & $c^1$ & $c^2$ & $c^3$ & $c^4$ & $c^5$\\
\hline
1&     1  &   1 &     1&     1 &    1\\
2&     1  &   2 &    2  &   2  &   2\\
3&     1  &   3 &   3   &  3   &  3\\
4&     1  &   4 &    4  &   4  &   4\\
5&     2  &   1 &    2  &   3  &   4\\
6&     2  &   2 &    1  &   4  &   3\\
7&     2  &   3 &    4  &  1   &  2\\
8&     2  &   4 &    3  &  2   &  1\\
9&     3  &   1 &    3  &   4  &   2\\
10&     3  &   2 &    4  &   3  &   1\\
11&     3  &   3 &    1  &   2  &   4\\
12&     3  &   4 &    2  &  1  &   3\\
13&     4  &   1 &    4  &  2  &   3\\
14&     4  &   2 &    3  &  1  &   4\\
15&     4  &   3 &    2  &  4  &   1\\
16&     4  &   4 &    1  &  3  &   2\\
\hline
\end{tabular}
\hspace{.4in}
\begin{tabular}{cccccc}
\hline
\multicolumn{6}{c}{An OA$(16,5,2,1)$} \\
\hline
Senarios\# & $c^1$ & $c^2$ & $c^3$ & $c^4$ & $c^5$\\
\hline
1&     1&     1&     1&     1&     1\\
2&     1&     2&     2&     2&     1\\
3&     2&     2&     2&     2&     2\\
4&     2&     1&     1&     1&     2\\
5&     2&     1&     1&     2&     2\\
6&     2&     2&     2&     1&     2\\
7&     1&     2&     2&     1&     1\\
8&     1&     1&     1&     2&     1\\
9&     1&     1&     2&     1&     2\\
10&     2&     2&     1&     1&     1\\
11&     2&     2&     1&     2&     1\\
12&     1&     1&     2&     2&     2\\
13&     2&     1&     2&     2&     1\\
14&     1&     2&     1&     2&     2\\
15&     1&     2&     1&     1&     2\\
16&     2&     1&     2&     1&     1\\
\hline
\end{tabular}
\label{tb:two_oa}
\end{table}

Comparing orthogonal arrays can be difficult.  As in \cite{Owen1994},
we define $\omega_{ij}(u)$ for each $u \subseteq \mathcal{D} :=
\{1,2,\ldots,m\}$ as $\omega_{ij}(u):=\{k \in u |
c_{ik}=c_{jk} \}$. We further define
\begin{equation}\label{eq:M}
M(u,r):=\sum_{i=1}^n \sum_{j=1}^n 1_{|\omega_{ij}(u)|=r},
\end{equation}
for $u \subseteq \mathcal{D}$ and $r=0,1,\ldots,|u|$ to be the
number of pairs of $i$th and $j$th rows in an orthogonal array $C$
($i$ can be the same as $j$) such that $c_i$ and $c_j$ agree on
exactly $r$ of the axes in $u$. For an OA$(N,m,s,2)$ without
coincidence defect, $M(u,3)=N$ for any $u \subseteq \mathcal{D}$. For
the array on the right in Table~\ref{tb:two_oa}, $M(u,3)$ may be much
larger than 16 --- for example, $M(\{1,2,3\},3)=3N$. In general, we
would like to select the orthogonal array $C$ with no coincidence
defect such that there are no duplicates in any three columns of
$C$. If we are forced to use orthogonal arrays with coincidence
defect, we should pick the one with the smallest value of
$M(u,r)$. Discussion on the existence of coincidence defects for
orthogonal arrays constructed using the methods of
Table~\ref{tb:oa_con} can be found in \cite{Owen1994}.

\subsection{Construction}\label{subsec:oaslh_con}

Let SOLH denote the scheme that generates batches $D_1,D_2,\dotsc,D_t$
as slices of a Latin hypercube design based on sliced orthogonal
arrays. The original purpose of SOLH was to share strength across all
batches for numerical integration with higher accuracy. Given an
OA$(N,m+1,t,2)$ with $N=nt$, $n=\lambda t$, and $s=t$ symbols, batches
$D_1,D_2,\dotsc,D_t$ each with $n$ scenarios can be constructed under SOLH
using Algorithm~\ref{alg:solh} \citep{Hwang2013}.

\begin{algorithm}
\caption{Generating a Sliced Orthogonal Array-Based Latin Hypercube Design}
\label{alg:solh}
\begin{itemize}
\item[] {\bf Step 1.}  Randomize the rows, columns and symbols of an
  OA$(N,m+1,t,2)$ to obtain an array $C=(c_{ik})_{N \times
    (m+1)}$. Let $C(:,1),C(:,2),\dotsc,C(:,m+1)$ denote $m+1$ columns
  of $C$.
\item[] {\bf Step 2.}  Rearrange the rows of $C$ so that
  $c_{i(m+1)}=\ell$ if $\left\lceil i/n \right\rceil=\ell$ for
  $\ell=1,2,\dotsc,t$. For $r=1,2,\dotsc,t$, let $A_r=(a_{r,ik})_{n
  \times m}$ denote a Latin hypercube design such that
  $a_{r,ik}=c_{[(r-1)n+i]k}$.

\item[] {\bf Step 3.}  For $r=1,2,\dotsc,t$ and $k=1,2,\dotsc,m$, do the
  following independently: replace all the $\ell$s in $A_r(:,k)$ by a
  uniform permutation on $Z_\lambda \oplus (l-1)t$ for
  $\ell=1,2,\dotsc,t$.

\item[] {\bf Step 4.}  Use $A_r$ thus obtained to construct $D_r$
  following steps 2 and 3 for SLH in \S~\ref{subsec:slhd_con}.
\end{itemize}
\end{algorithm}

We obtain a sliced orthogonal array based Latin hypercube design by
vertically stacking $D_1,D_2,\dotsc,D_t$. The construction above
exploits the fact that taking the scenarios in an OA$(N,m+1,s,\tau)$ that
have the same level in the first column, and deleting that first
column, gives an OA$(N/s,m,s,\tau)$ \citep{Hedayat1999}.

\begin{figure}[!h] \centering
	\includegraphics[scale=.7]{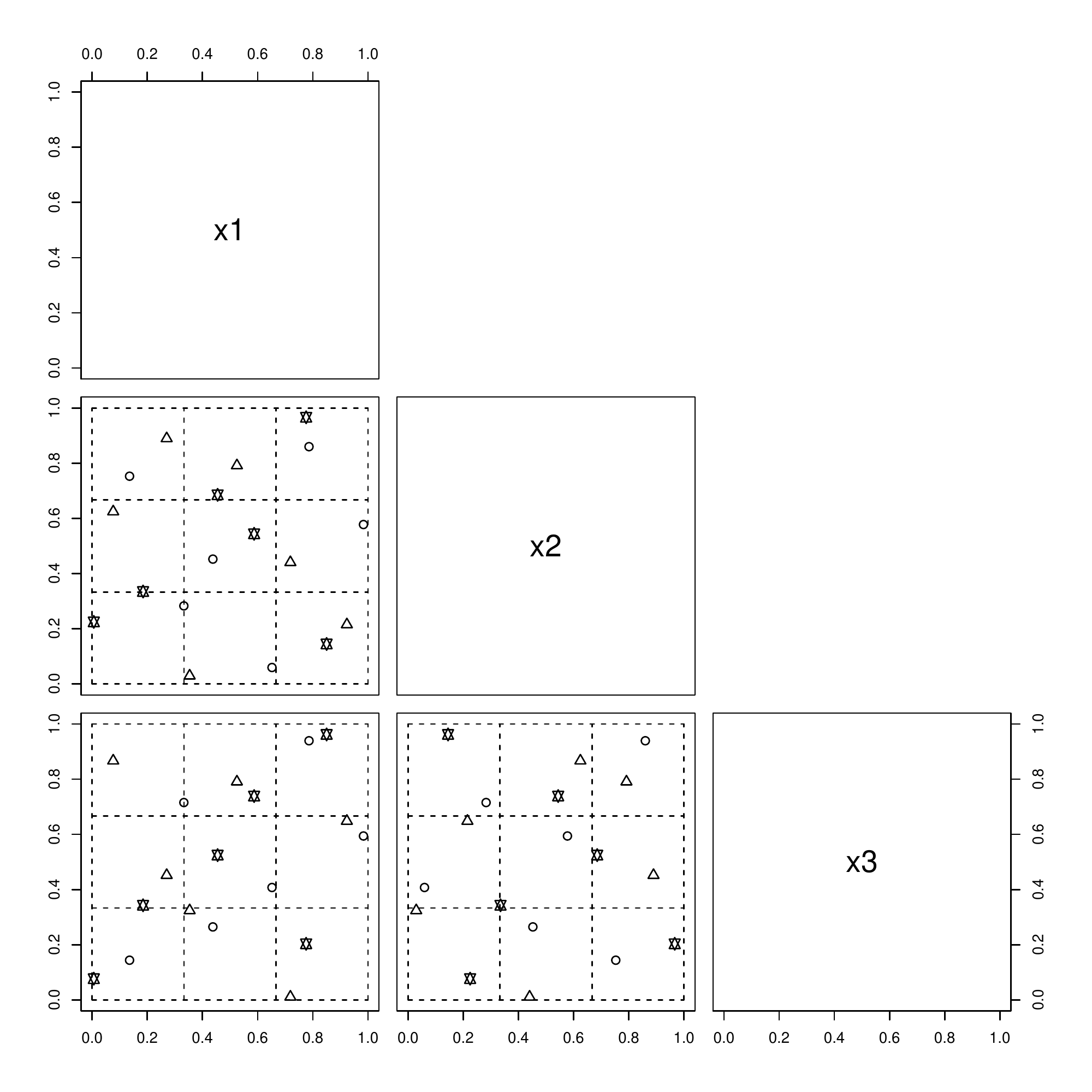}
	\caption{Bivariate projections of a sliced orthogonal array
          based Latin hypercube design $\mathbf{D}$ with batches
          $D_1$ (circles), $D_2$ (triangles), and
          $D_3$ (stars).}
	\label{fig:oslhd}
\end{figure}

Figure~\ref{fig:oslhd} presents bivariate projections of a Latin
hypercube design based on sliced orthogonal arrays, with batches
$D_1$, $D_2$, $D_3$ based on an OA$(18,4,3,2)$. For any $D_r$, each of
the six equally spaced intervals of $(0,1]$ contains exactly one scenario
  in each dimension. When combined, each of the 18 equally spaced
  intervals of $(0,1]$ contains exactly one scenario. Additionally, each of
    the $3 \times 3$ squares in the dashed lines has exactly two scenarios,
    because $\lambda=2$.

\subsection{Theoretical Results}\label{subsec:oaslh_thm}

We provide a general variance formula of $L_{n,t}$ under SOLH
introduced in the last section. The result is a direct consequence of
\cite[Theorem~1]{Hwang2013}.

\begin{theorem}\label{thm:var_oslhd}
Consider problem \eqref{eq:two_stage}, and suppose that the conditions
in Proposition~\ref{prop:equal} hold. Based on an OA$(N,m+1,t,2)$ with
$N=nt$, $n=\lambda t$, and $s=t$ symbols, we have as $s \rightarrow
\infty$ that
\begin{equation*}
\var(L_{n,t}^{SOLH})=N^{-2} \sum_{|u| \geq 3} M(u,|u|)\var[\tilde{G}_u(x^*,\tilde\xi^u)]+o(N^{-1}),
\end{equation*}
where $u$ is a subset of $\mathcal{D}$, and $\tilde\xi^{u}$ contains $\tilde\xi^{k}$ for $k \in
u$.  
\end{theorem} 

\proof  \cite[Theorem~1]{Hwang2013} proves this result
for continuous $\tilde{G}(x^*,\cdot)$. We extend it to cases in which
$\tilde{G}(x^*,\cdot)$ is a step function (since $\xi$ has finite
support) by applying \cite[Lemma~3]{Loh1996}. \Halmos \endproof

Theorem~\ref{thm:var_oslhd} shows that the variance of $L_{n,t}$ can
be reduced under SOLH after filtering out the bivariate interactions
in $\tilde{G}(x^*,\tilde\xi)$. This fact remains true for SOLH based
on any strength-two orthogonal array even if coincidence defects are
present, provided that $s \rightarrow \infty$. According to the
functional ANOVA decomposition and Theorem~\ref{thm:var_slhd} (i), we
have $\var(L_{n,t}^{SLH})=N^{-1} \sum_{|u| \geq 2}
\var[\tilde{G}_u(x^*,\tilde\xi^u)]+o(N^{-1})$, where coefficients of
variances due to interactions with more than two variables are all of
order $N^{-1}$. As a result, SOLH based on an orthogonal array with
$M(u,|u|)>N$, for some $u \subseteq \mathcal{D}$ and $|u| \geq 3$, can
inflate the variance due to higher-order interactions. This side
effect is less significant if the bivariate interactions dominate
higher order interactions, which is true in most problems.

In addition, we need to emphasize that each batch under SOLH is based
on an OA$(n,m,t,1)$, which is a Latin hypercube design instead of an
ordinary Latin hypercube. We conjecture that
Assumption~\ref{ass:consist} still holds in this case, and the
bivariate variance reduction is due to the between-batch negative
dependence rather than the within-batch non-ordinary Latin hypercube
structure. This conjecture is consistent with our computational
observations, as shown below. 

\subsection{Practical Considerations}\label{subsec:oaslh_prac}

Using the SOLH method is not just a matter of picking the desired $n,m$
values of the dimensions of the design matrices and the number of
batches $t$, like it was in the SLH case. Picking the appropriate
orthogonal array requires some understanding of the tradeoffs between
the different choices. 

Consider an OA$(n^2,m+1,n,2)$. The benefits of using these arrays
include (i) each batch is based on an ordinary Latin hypercube, and
(ii) there is no coincidence defect. However, we have to solve $n$
batches of $n$ scenarios each. We want $n$ to be large enough to ensure that
the quantity $x_n^*$ converges and that the bias
$v^*-\mathbb{E}[v_n(D)]$ is small, but solving $n$ batches could be
computationally infeasible. 

One way around this computational hurdle is to relax the restriction
that we use all the batches that are generated by the SOLH method. We
can use just the first $t$ batches, and we will call this the sliced
partial orthogonal-aray-based Latin hypercube method (SPOLH). Let
$\upsilon$ denote the fraction of the number of batches generated by
SOLH. SPOLH can perform better than SLH since $\upsilon$ of the
variance due to bivariate interactions can be removed. However, when
$\upsilon$ is small (10\%, say), it does not guarantee substantial
variance reduction from SLH to SPOLH.
 
We can alternatively use orthogonal arrays with a higher index value
(i.e.\ $\lambda > 1$) allowing  us to have a $t$ that is much smaller than
$n$.  In this case, there is coincidence defect and picking and generating the
right orthogonal arrays with desirable $n,m,t$ values with low $M(u,r)$
can be difficult.

\section{Experimental Setup} \label{sec:expsetup}

To illustrate the effectiveness of negatively dependent designs for
estimation of the lower bound on the objective value in stochastic
programs, we perform computational experiments using data sets from
the literature. This section provides some details on our
implementations and test problems.

\subsection{Implementation} \label{subsec:implemenation}

We estimate a single lower bound by choosing a design family,
obtaining a set of sampled approximations using this design family,
and finally solving these approximations. The performance of a
sampling scheme is assessed by repeating this process to obtain a
number of lower bound estimates.  We then calculate the mean and
variance of these estimates.

We augmented the SUTIL library \citep{SUTIL}, a C and C++ based
library for manipulating stochastic programming instances with a
design-based sampling framework.  This augmented SUTIL library can
take an orthogonal array as input and generate an
orthogonal-array-based design family as output. The library can
generate other design families from scratch. We used a C library due
to Owen (available at \url{http://lib.stat.cmu.edu/}) to generate the
orthogonal arrays.  The SUTIL software produces the extensive form (or
deterministic equivalent) linear program of each sampled instance and
outputs the linear program as an MPS file.  These files are fed into
the commercial linear programming solver Cplex v12.5 and solved using
the barrier method.  All the timed experiments were run on an Intel
Xeon X5650 (24 cores @ 2.66Ghz) server with 128GB of RAM, and 16 of
the 24 cores available on the machine were used.  Other experiments
were run on the HTCondor grid \citep{condor-practice} of the Computer
Sciences department at University of Wisconsin-Madison. They required
more than a week of wall-clock time to complete.

\subsection{Test Problems}

Our five test problems were drawn from the stochastic programming
literature. In fact, they are the same problems that were studied in
\cite{Linderoth2006}. These problems are specified in SMPS file format
(a stochastic-programming extension of MPS), and have finite discrete
distributions for their random variables.

\begin{itemize}
\item[] {\tt 20term}, first described in \cite{Mak1999}, is a model of
  a motor freight carrier's operations. The first stage variables
  represent positions of a fleet of vehicles at the beginning of a
  day. The second-stage variables determine the movements of the
  fleet through a network to satisfy point-to-point demands for
  shipments (with penalties for unsatisfied demands), and to end the
  day with the fleet back in their initial positions.

\item[] {\tt gbd} is derived from an aircraft allocation problem
  originally described in the textbook of
  \citet[Chapter~28]{Dantzig1963}. Four different types of aircraft are
  to be allocated to routes to maximize the profit under uncertain
  demand for each route. There are costs associated with using each
  aircraft type, and when the capacity does not meet demand.

\item[] {\tt LandS} is a modification by \cite{Linderoth2006} of a
  simple problem in electrical investment planning described by
  \cite{Louveaux1988}. The first-stage variables represent capacities
  of different new technologies, and the second-stage variables
  represent production of each of the different modes of electricity
  from each of the technologies.

\item[] {\tt ssn} \citep{Sen1994} is a problem from telecommunications
  network design.  The owner of a network sells private-line services
  between pairs of nodes in the network. During the first stage of the
  problem, the owner has a budget for adding bandwidth (capacity) to
  the edges in the network.  In the second stage, to satisfy uncertain
  demands for service between each pair of nodes, short routes between
  two nodes with sufficient total bandwidth must be identified. Unmet
  demands incur costs, and the goal of the problem is to minimize the
  expected costs.

\item[] {\tt storm} is a cargo flight scheduling problem described by
  \cite{Mulvey1995}. The goal is to schedule cargo-carrying flights
  over a set of routes in a network, where the amount of cargo
  delivered to each node is uncertain. In the first stage, the number
  of planes of each type on each route is decided. In the second
  stage, the random variables are the demands on the amounts of cargo
  to be delivered between nodes. The goal is to minimize (1) the costs
  that comes from assigning the planes and balancing payloads, and (2)
  the penalties associated with unmet demands.
\end{itemize}

All of these problems fit our monotonicity assumption defined in \S~\ref{subsec:samp_slhd}. The
intuition for this property is straightforward -- the objective in each
problem is to minimize costs and penalties that come from unmet
demands. Hence, as the demand increases, the objective value always
increases.

We outline some facts about the random variables and optimal solution
for each problem in Table~\ref{tb:spdetails}. The random variables
for all the problems are independent from each other. For the 
distribution column, we use `uniform' to refer to a uniform 
distribution over all possible values, and `irregular' for any other
distribution. Additionally, we note that the distribution of each of
the random variables of {\tt LandS} approximate a linear function as it
has a uniform distribution over evenly spaced points. The
upper- and lower-bound estimates (with 95\% confidence intervals) are
obtained from \cite[Table~4]{Linderoth2006} with $n = 5000$. We will
refer to the quantities in this table in our discussion of the
computational results.

\begin{table}[bhtp]
    \centering
    \caption{Properties of our stochastic programming test problems}
    {\small
    \begin{tabular}{lcccrclrcl}
        \hline
        &\multicolumn{3}{c}{Random Variables}&\multicolumn{6}{c}{Bounds (95\% confidence interval)} \\
        &Number&Possible Values&Distribution&\multicolumn{3}{c}{Lower}&\multicolumn{3}{c}{Upper}\\
        \hline
        20term&40&2&Uniform&$254298.57$&$\pm$&$38.74$&$254311.55$&$\pm$&$5.56$  \\
        gbd&5&13 - 17&Irregular&$1655.62$&$\pm$&$0.00$&$1655.628$&$\pm$&$0.00$\\
        LandS&3&100&Uniform&$ 225.62$&$\pm$&$0.02 $&$ 225.624$&$\pm$&$0.005 $\\
        ssn&86&4 - 7&Irregular&$ 9.84$&$\pm$&$ 0.10 $&$ 9.913 $&$\pm$&$ 0.022 $\\
        storm&117&5&Uniform&$ 15498657.8$&$\pm$&$ 73.9 $&$ 15498739.41 $&$\pm$&$ 19.11 $\\
        \hline
    \end{tabular}
    }
    
    \label{tb:spdetails}
\end{table}

\section{Computational Results}\label{sec:compute}

In this section we will summarize the computational results and
observations.  For each combination of sampling scheme, number of
scenarios, and number of batches, we perform the lower bound
estimation process 100 times, and compute the mean and standard error
of the 100 lower bounds obtained.

Before we proceed with our observations, we should emphasize that
while we use the same problems as in \cite{Linderoth2006}, our results
are not directly comparable. In particular, \cite{Linderoth2006} focus
on results based on analysis of the objective values across 10
uncorrelated subproblems of a single replicate (or similarly, across
10 replicates of one subproblem each).  We focus on the analysis
of the mean SAA objective values between many (potentially correlated)
subproblems across 100 replicates.  Hence, the results in
\cite{Linderoth2006} yield a measure of the quality of a single
lower-bound estimate, while our results compare the quality of several
different lower bound estimates.

All data used in this analysis --- specifically, the objective values
obtained by solving each LP corresponding to each batch --- can be
obtained at the web site for this paper at
\url{http://pages.cs.wisc.edu/~conghan/slhd/}.

\subsection{Timing Information} \label{subsec:timing}

Table~\ref{tb:exptime} shows the wall-clock time required to solve a
single sampled approximation for the five test problems using our
timing setup as described in \S~\ref{subsec:implemenation}. Even though the time
to generate a batch of scenarios increases as the sampling method
increases in complexity, this time is insignificant compared to the
rest of the process and hence was not included in this table. The
timings were obtained for each problem and number of scenarios by
averaging across $16$ subproblems constructed from all design methods
we considered. The timing increases with the number of scenarios at a 
superlinear rate. Hence, when the underlying problem is difficult
and the number of scenarios is sufficiently high, solving many batches 
could be extremely time consuming even with multiple machines.

\begin{table}[bhtp]
    \centering
    \caption{Wall clock times (in seconds) for solving one sample
      approximation problem with varying numbers of scenarios}
    \label{tb:exptime}
    \begin{tabular}{|l|rrr|}
        \hline
& \multicolumn{3}{c|}{no. of scenarios} \\
        &256 &512 &1024 \\
        \hline
        20term&4.83&11.12&30.09\\
        gbd&0.03&0.10&0.17\\
        LandS&0.07&0.12&0.25\\
        ssn&10.36&23.80&61.63\\
        storm&16.40&37.85&84.22\\
        \hline
    \end{tabular}
\end{table}

\subsection{Mean and Standard Error of the Different Sampling Methods}

In our first set of experiments, we computed the means and standard
errors of the lower bound estimates of the objective value over 100
replicates of each sampling method and each number of scenarios. We
fixed the number of subproblems at $t=16$, while varying the number of
scenarios per batch according to $n \in \{ 128, 256, 512, 1024\}$.  We
tested four different sampling methods: Monte Carlo (MC), independent
ordinary Latin hypercube (ILH), Sliced Latin hypercube (SLH) and
Sliced Partial Orthogonal Array Based Latin hypercube based on Bush
orthogonal arrays (BUSH), which are $OA(n^2,m+1,n,2)$.  Since we are
only using 16 out of $n$ possible batches, we are discarding much of
the orthogonal array, and therefore not achieving much two-dimensional
stratification in our negatively dependent designs. Computational
results are summarized in Table~\ref{tb:expone}.

\begin{table}[bhtp]
    \centering
    \caption{Mean and variance of estimates of the lower bound with 16
      batches, over 100 replicates.}
    \label{tb:expone}
    {\small
    \begin{tabular}{llcccccccc}
        \hline
        && \multicolumn{2}{c}{128 scenarios} &
           \multicolumn{2}{c}{256 scenarios} &
           \multicolumn{2}{c}{512 scenarios} &
           \multicolumn{2}{c}{1024 scenarios} \\
        &&Mean&SE&Mean&SE&Mean&SE&Mean&SE\\
        \hline
        \multirow{4}{*}{20term}&MC&254253.1&244.1&254292.1&160.2&254297.4&95.1&254311.3&69.3\\
        &ILH&254296.7&58.2&254311.1&39.8&254306.4&25.6&254311.0&19.9\\
        &SLH&254285.9&53.3&254303.0&45.0&254307.6&31.5&254310.0&23.8\\
        &BUSH&254296.0&54.8&254299.2&38.3&254307.5&30.2&254312.8&22.1\\
        \hline
        \multirow{4}{*}{gbd}&MC&1653.207&14.292&1654.061&10.540&1655.124&6.881&1656.837&5.585\\
        &ILH&1655.637&1.094&1655.760&0.670&1655.606&0.260&1655.616&0.157\\
        &SLH&1655.640&0.281&1655.622&0.171&1655.615&0.066&1655.635&0.044\\
        &BUSH&1655.602&0.275&1655.609&0.167&1655.645&0.080&1655.632&0.043\\
        \hline
        \multirow{4}{*}{LandS}&MC&225.3951&1.2923&225.7044&0.9876&225.5539&0.6106&225.6881&0.4622\\
        &ILH&225.6314&0.0549&225.6233&0.0332&225.6249&0.0278&225.6301&0.0158\\
        &SLH&225.6135&0.0471&225.6248&0.0370&225.6259&0.0255&225.6295&0.0177\\
        &BUSH&225.6159&0.0522&225.6191&0.0319&225.6274&0.0266&225.6295&0.0185\\
        \hline
        \multirow{4}{*}{ssn}&MC&7.426&0.387&8.403&0.287&9.028&0.188&9.411&0.142\\
        &ILH&8.945&0.267&9.378&0.186&9.635&0.150&9.770&0.103\\
        &SLH&8.877&0.225&9.321&0.203&9.609&0.137&9.775&0.087\\
        &BUSH&8.929&0.258&9.374&0.181&9.656&0.134&9.785&0.091\\
        \hline
        \multirow{4}{*}{storm}&MC&15498662.4&7441.1&15498518.5&5517.1&15498532.9&3648.5&15498473.7&2838.3\\
        &ILH&15498741.0&454.9&15498678.8&257.0&15498698.9&151.2&15498716.6&98.2\\
        &SLH&15498690.0&245.2&15498683.9&159.0&15498695.4&104.6&15498721.5&79.3\\
        &BUSH&15498699.6&238.7&15498731.9&149.6&15498688.7&115.6&15498709.3&85.4\\
        \hline
        \end{tabular}
    }
\end{table}

We begin with some general observations about the mean of the lower
bound estimates.  As observed in other experiments
\citep{Linderoth2006,Freimer2012}, the Monte Carlo method is
significantly worse than Latin Hypercube-based methods in terms of the
bias.  As expected, the Latin Hypercube-based methods produce
statistically-indistinguishable lower bounds.

By comparing with the results of Table~\ref{tb:spdetails}, for all
problems except {\tt ssn}, ILH attains a mean extremely close to the
true mean when the number of scenarios is 1024, and is already very
close with a smaller number of scenarios. {\tt ssn} is known to be a
challenging problem, requiring at least 512 scenarios to attain a
reasonable estimate of the optimum even for the three Latin hypercube
schemes.  Increasing the number of scenarios beyond 1024 would
continue to improve the quality of the estimates. \cite{Linderoth2006}
provide a more detailed description of the behavior of SAA for the
{\tt ssn} problem.

We now turn to the standard error. By this measure, for every problem
except {\tt ssn}, we see that MC performs much worse than the other
methods. It is only slightly worse for {\tt ssn}. This table shows
situations in which the negatively dependent designs of SLH and BUSH
begin to distinguish themselves from ILH. For {\tt gbd} and {\tt
  storm}, we can see large improvements from ILH to SLH/BUSH. The
improvement from ILH to SLH/BUSH is much less pronounced in {\tt ssn},
being in general smaller than the improvement from MC to ILH, and in
one case performing slightly worse than ILH.  However, we should note
that in all the problems, BUSH and SLH have roughly similar
performance. This observation suggests that when we use too few
subproblems, the effect of partial orthogonality on performance is not
significant.

For {\tt 20term} and {\tt LandS}, the Latin hypercube schemes ILH, SLH,
and BUSH perform similarly.  In the case of {\tt 20term}, this
similarity is unsurprising.  The benefits of SLH over ILH come from the
increased stratification that comes from the sliced structure, but
since each variable can only take on two values, each with probability
$0.5$, any stratification that divides the probability space into a
multiple of two would perform equally well. In the case of {\tt LandS},
we suspect that the similarity of performance is due to the smoothness
of the distribution of the random variables. In fact, the cumulative
distribution function is essentially linear.  We have found that when
we modify the distribution to be more irregular or to to be a
uniform distribution over a much smaller set, it tends to drive up the standard error and to cause SLH to have a significantly smaller standard error than ILH. 

We now consider a greater number of subproblems for each $n$ and new
alternatives for the underlying orthogonal arrays. In addition to the
four different sampling methods considered earlier, we show two
additional variants: Sliced Orthogonal Array Based Latin hypercube
based on Bose-Bush orthogonal arrays (BB), and independent batches
taken over several BB (INDBB).  Bose-Bush orthogonal arrays have an
OA$(\lambda s^2, m+1,s,2)$ structure.  We pick $s$ to be equal to the
number of subproblems, and define $\lambda = n/s$. With these choices,
the sliced designs achieve full two-dimensional stratification, making
BB an example of SOLH sampling.

We include INDBB in our experiments to help isolate the factors that
lead to the stronger performance of BB. If each slice of the BB design
has some special structure that leads to improved performance, then
INDBB designs should perform better than ILH, and the performance of
INDBB should be close to that of BB.  However, if the performance gains
of BB are primarily due to the better two-dimensional stratification,
then we would expect INDBB to perform no better than ILH. The numerical
results support the second claim.  

\begin{table}[bhtp]
    \centering
    \caption{Mean and standard error of estimates of the lower bound with 32 or
             64 batches (over 100 replicates)}
    \label{tb:exptwo}
    {\small
    \begin{tabular}{llcccccccc}
        \hline
\multicolumn{2}{c}{(scenarios, batches)} & 
           \multicolumn{2}{c}{(128,32)} &
           \multicolumn{2}{c}{(256,32)} &
           \multicolumn{2}{c}{(512,64)} &
           \multicolumn{2}{c}{(1024,64)} \\
\hline
        &&Mean&SE&Mean&SE&Mean&SE&Mean&SE\\
        \hline
        \multirow{6}{*}{20term}&MC&254295.4&164.5&254305.0&116.3&254301.1&54.4&254313.4&36.9\\
        &ILH&254305.7&43.4&254296.4&29.2&254307.0&16.8&254309.1&9.9\\
        &SLH&254294.2&45.8&254306.3&28.5&254306.2&14.3&254308.5&11.3\\
        &BUSH&254293.7&36.5&254305.3&27.8&254306.7&13.6&254309.4&10.1\\
        &BB&254296.1&20.9&254305.3&18.9&254307.3&7.6&254310.3&4.9\\
        &INDBB&254294.4&39.3&254301.1&29.9&254308.3&15.2&254309.1&11.1\\
        \hline
        \multirow{6}{*}{gbd}&MC&1653.130&9.485&1654.699&7.091&1655.488&3.389&1655.655&2.986\\
        &ILH&1655.550&0.849&1655.658&0.390&1655.633&0.164&1655.628&0.094\\
        &SLH&1655.649&0.169&1655.620&0.066&1655.628&0.017&1655.628&0.011\\
        &BUSH&1655.628&0.163&1655.628&0.066&1655.629&0.017&1655.628&0.011\\
        &BB&1655.614&0.170&1655.626&0.072&1655.629&0.018&1655.627&0.011\\
        &INDBB&1655.618&0.799&1655.582&0.483&1655.618&0.140&1655.641&0.096\\
        \hline
        \multirow{6}{*}{LandS}&MC&225.6448&0.9108&225.6300&0.6092&225.6431&0.2844&225.5845&0.2202\\
        &ILH&225.6151&0.0344&225.6190&0.0248&225.6255&0.0131&225.6298&0.0088\\
        &SLH&225.6172&0.0351&225.6260&0.0263&225.6253&0.0124&225.6287&0.0087\\
        &BUSH&225.6155&0.0332&225.6247&0.0225&225.6260&0.0116&225.6269&0.0081\\
        &BB&225.6178&0.0068&225.6247&0.0047&225.6270&0.0015&225.6282&0.0010\\
        &INDBB&225.6202&0.0370&225.6220&0.0252&225.6258&0.0123&225.6281&0.0085\\
        \hline
        \multirow{6}{*}{ssn}&MC&7.426&0.275&8.412&0.197&9.011&0.094&9.389&0.071\\
        &ILH&8.908&0.184&9.378&0.127&9.628&0.073&9.767&0.045\\
        &SLH&8.911&0.198&9.386&0.131&9.614&0.064&9.771&0.054\\
        &BUSH&8.905&0.175&9.390&0.123&9.634&0.064&9.770&0.051\\
        &BB&8.925&0.105&9.408&0.083&9.627&0.029&9.767&0.022\\
        &INDBB&8.938&0.185&9.381&0.134&9.620&0.078&9.763&0.049\\
        \hline
        \multirow{6}{*}{storm}&MC&15499036.3&5185.4&15498564.6&3653.0&15498659.9&2039.3&15498513.0&1263.0\\
        &ILH&15498674.5&377.4&15498717.2&179.3&15498690.9&68.6&15498715.7&43.3\\
        &SLH&15498658.3&149.0&15498712.7&99.8&15498699.4&53.4&15498718.1&37.0\\
        &BUSH&15498687.3&133.0&15498707.0&104.1&15498701.9&47.0&15498721.9&38.4\\
        &BB&15498686.1&76.1&15498710.4&42.5&15498693.4&22.0&15498720.7&12.6\\
        &INDBB&15498674.6&297.4&15498712.5&201.2&15498695.8&67.9&15498720.7&43.1\\
        \hline
    \end{tabular}
    }
\end{table}

Results are shown in Table~\ref{tb:exptwo}. Many of the observations
about MC/ILH/SLH/BUSH from Table~\ref{tb:expone} carry over. Also,
comparing the standard error of MC/ILH/SLH/BUSH between the two tables,
we notice a factor of $\sqrt{2}$ or $2$ difference in standard error,
depending on whether the number of subproblems was doubled or
quadrupled. This factor is consistent with Theorem \ref{thm:large_t}.
The lower bound estimates are roughly the same in both tables,
demonstrating that changing the number of subproblems does not affect
the bias.

Table~\ref{tb:exptwo} shows a considerable advantage for BB over ILH.
In fact, BB is better than all other methods tested, except on {\tt
  gbd}, where it performs similarly to SLH/BUSH.  On {\tt LandS}, BB
performs about 5-10$\times$ better than the other sliced sampling
methods. A possible explanation for this huge improvement is that the
total number of random variables is just three, so having
two-dimensional stratification would cover a large portion of the
possible interactions between variables. In the case of {\tt ssn}, the
improvement from ILH to BB is comparable to the improvement from MC to
ILH, a bigger factor than is observed for any other problem.

Finally, we turn our attention to running times. The standard error of
the estimates between the results for 512 scenarios and 16 slices in
Table~\ref{tb:expone} and 256 scenarios and 32 slices in
Table~\ref{tb:exptwo} are similar. Since the timing scales
superlinearly in the number of scenarios (\S~\ref{subsec:timing}), the
amount of time it takes to solve $c t$ sampled approximations of $n$
scenarios sequentially can be substantially less than solving $t$
sampled approximations of $c n$. Each batch could also be solved 
independently and in parallel. This suggests that if computing 
resources on each machine is limited and using SLH/SPOLH with $c n$ 
scenarios and $t$ batches is computationally infeasible, using SOLH 
with with $n$ scenarios and $c t$ batches can be an effective way of 
reducing the standard error.

We conclude that for a fairly small number of subproblems, the sliced
sampling methods perform at least as well as Latin hypercube sampling,
and in fact show significant improvement in some cases. Once we
increase the number of subproblems and exploit the full
``orthogonality'' property of the orthogonal arrays, we see a
substantial improvement in {\em all} cases.  Thus, if the
computational budget will only allow a small number of batches for the
given value of $n$ for which a lower bound $v_n$ is being estimated,
there is significant computational benefit to using the more
sophisticated sampling methods introduced in this work.

\section{Conclusions and Future Work}\label{sec:dis}

In this paper, we propose the use of two types of negatively dependent
designs to improve the lower bound of the objective value. Sliced
Latin hypercube sampling is easy to implement since SLH does not
impose any restriction on the number of batches $t$ and the number of
scenarios in each batch. We introduce the concept of monotonicity for
two-stage stochastic linear programs, and we provide a non-asymptotic
result showing that SLH can be better than ILH for problems with this
monotonicity property. On the other hand, we show that SLH is
asymptotically equivalent to ILH if the distribution of the random
vector has finite support and the approximate solutions converge. Our
computational results supports the theory, showing that SLH performs
no worse than ILH and in some cases performs significantly better than
ILH.

To improve upon SLH, we consider sliced orthogonal array-based Latin
hypercube sampling schemes, which achieve stronger negative dependence
between batches. The choice of the underlying orthogonal array can
make a huge difference in variance reduction. We provide empirical
results showing that when we are able to exploit the full
orthogonality of the underlying orthogonal array, using Bose-Bush
orthogonal arrays \citep{BoseBush1952}, the performance is
significantly better than when we use only part of an orthogonal
array.

Our work treats Latin hypercube sampling as the baseline method, in
part because it was investigated in earlier work
\citep{Linderoth2006}. Other sampling methods, such as $U$ sampling \citep{Tang1993,Tang2010} and
randomized quasi-Monte Carlo
\citep{Niederreiter:1992,Owen:1995,Mello2008} could have been used
instead for constructing a single SAA problem. We can carry over the idea
of negatively dependent designs to these advanced within-batch
sampling techniques. For $U$ sampling, a strength-two orthogonal
array-based Latin hypercube design could be generated for each SAA
problem. The $t$ underlying strength-two orthogonal arrays can be
obtained by slicing a larger strength-three orthogonal
array. For randomized quasi-Monte Carlo, we can sample different batches based
on the same low-discrepancy sequence such that batches are negatively dependent spontaneously. Comparing with Latin hypercube sampling, both $U$ sampling and randomized quasi-Monte Carlo are extremely
complicated to implement as they require more constraints on the selection of batch size $n$ and the number of batches $t$. A potential research direction in the future for us is to study the theoretical and empirical performance of negatively dependent batches based on these advanced sampling methods.

\bibliographystyle{ormsv080}
\bibliography{sp}

\end{document}